\documentclass[aps,prl,amsmath,amssymb,preprint]{revtex4-2}

\usepackage{float}
\usepackage{graphicx}
\usepackage{amsmath}
\usepackage{amssymb}
\usepackage{color}
\usepackage{soul}
\usepackage{subfigure}
\usepackage{stackengine}
\usepackage{siunitx}

\usepackage{mathrsfs}
\newcommand{\figScwPL}{S4}
\newcommand{\figScwRaman}{S5}
\newcommand{\figSPLtwoone}{S6}
\newcommand{\figSHOTPL}{S7}
\newcommand{\figSfitHOTPL}{S8}
\newcommand{\figSTeneh}{S9}
\newcommand{\HeteroSpectrum}{S10}
\newcommand{\HeteroSpectrumone}{S11}
\newcommand{\figSGgr}{S12}

\begin{document}

\title{Picosecond energy transfer in a transition metal dichalcogenide-graphene heterostructure revealed by transient Raman spectroscopy}

\author{%
	Carino Ferrante$^{1,2,3,4\ddagger}$,
	Giorgio Di Battista$^{4,5,\ddagger}$,
	Luis E. Parra L\'{o}pez$^{5}$,
	Giovanni Batignani$^{4,3}$,
	Etienne Lorchat$^{5}$,
	Alessandra Virga$^{4,3}$,
	St\'ephane Berciaud$^{5,\dagger}$ and
	Tullio Scopigno$^{4,1,3,\dagger}$}

\affiliation{$^{1}$ Istituto Italiano di Tecnologia, Graphene Labs, Via Morego 30, I-16163 Genova, Italy}
\affiliation{$^{2}$  FSN-FISS-SNI Laboratory, ENEA, Casaccia R.C. Via Anguillarese 301, 00123 Roma, Italy}
\affiliation{$^{3}$ 	Center for Life Nano Science @Sapienza, Istituto Italiano di Tecnologia, Viale Regina Elena 291, I-00161, Roma,Italy}
\affiliation{$^4$ Dipartimento di
	Fisica,~Universit\`a~di~Roma~``La Sapienza", Piazzale Aldo Moro 5,~00185,~Roma,~Italy}
\affiliation{$^5$ Universit\'e de Strasbourg, CNRS, Institut de Physique et Chimie des Mat\'eriaux de Strasbourg (IPCMS), UMR 7504, F-67000 Strasbourg, France}
\affiliation{$\ddagger$ These authors contributed equally to this work}
\affiliation{$\dagger$ email: stephane.berciaud@ipcms.unistra.fr, tullio.scopigno@uniroma1.it }

\begin{abstract}
	Intense light–matter interactions and unique structural and electrical properties make Van der Waals heterostructures composed by Graphene (Gr) and monolayer transition metal dichalcogenides (TMD) promising building blocks for tunnelling transistors, flexible electronics,	 as well as optoelectronic devices, including photodetectors, photovoltaics and quantum light emitting devices (QLEDs), bright and narrow-line emitters using minimal amounts of active absorber material.
	The performance of such devices is critically ruled by interlayer interactions which are still poorly understood in many respects. Specifically, two classes of coupling mechanisms have been proposed: charge transfer (CT) and energy transfer (ET), but their relative efficiency and the underlying physics is an open question.
	Here, building on a time resolved Raman scattering experiment, we determine the electronic temperature profile of Gr in response to TMD photo-excitation, tracking the picosecond dynamics of the G and 2D bands. Compelling evidence for a dominant role ET process accomplished within a characteristic time of $\sim 4$ ps is provided.
	Our results suggest the existence of an intermediate process between the observed picosecond ET and the generation of a net charge underlying the slower electric signals detected in optoelectronic applications.
\end{abstract}

\maketitle

\section*{Introduction}Van der Waals (vdW) heterostructures have recently emerged as a versatile platform to combine, in a single unit, key properties of layered materials. For instance, TMD-Gr heterostructures have been thoroughly studied\cite{georgiou2013vertical,Britnell1311,qled} as they represent a new class of truly two-dimensional metal-semiconductor junctions useful for engineering broadband, efficient and ultrafast photodectors \cite{massicotte,massicotte2016} as well as narrow-line light-emitters \cite{berciaud_2020}. These systems jointly exploit the rich TMD photophysics \cite{Wang2018,trovatello2021optical,pogna2016photo,malard} and their high photodetection efficiency \cite{Mak2016} together with the unique electronic properties and sub-picosecond photoresponse of graphene (Gr) \cite{Tielrooij2013,koppens}. 

Remarkably, vdW heterostructures feature atomically sharp heterointerfaces whereat efficient charge tunnelling and/or energy funneling from the TMD to Gr layers can be achieved \cite{Froehlicher2018,Yuan2018,massicotte,He2014,Chen2019}.
While a solid ground has been established to describe such near-field couplings in nanoscale systems \cite{Govorov2016,Adams2003}, the ultimate thinness of two-dimensional materials as well as their particularly strong Coulomb and light-matter interactions lead to new regimes where the shares of charge and energy transfer and the associated timescales are challenging to determine \cite{Froehlicher2018,Kozawa2016}.
In order to conceive high-performance optoelectronic devices based on vdW heterostructures, a key challenge consists in spatially and temporally tracking charge carriers and excitons in tightly coupled layers. Along these lines, early measurements on TMD-Gr heterostructures have employed transient absorption (TA) spectroscopy \cite{He2014} and opto-electronic studies based on photoconductivity and photocurrent measurements \cite{massicotte}. TA spectroscopy benefits from a short time resolution, ultimately limited by the laser pulse duration. Although accelerated picosecond transient dynamics of TMD excitons in TMD-Gr heterostructures has been observed \cite{He2014,Yuan2018,Chen2019,Froehlicher2018,Aeschlimann2020,krause2021microscopic,zhou2021deciphering,Fu2020}, TA does not make it possible to disentangle contributions from net charge tunnelling and energy transfer, both when the latter arises from long-range dipole-dipole (F\"orster energy transfer) \cite{Selig2019} or short-range exchange-mediated (Dexter energy transfer) \cite{dexter} interactions.
In electrically contacted devices, photo-induced charge transfer from the TMD to Gr leads to  photogating \cite{Zhang2014}.
Such photogating processes \cite{Froehlicher2018,Lin2019,Miller2015,Furchi2014b} are environment-dependent and slow, involving charge transfer to Gr and subsequent back-transfer to the TMD monolayer on timescales ranging from ns to seconds \cite{Ahmed2020}. Alternatively, direct photocurrent generation is observed on faster (sub-nanosecond) timescales in vertically-biased  Gr/few-layer TMD/Gr junctions. This mechanism, however becomes quite inefficient when the active TMD region is thinned down to a monolayer  \cite{massicotte}, likely due to picosecond non-radiative transfer to graphene.

\begin{figure}
	\centering
	\includegraphics[width=14cm]{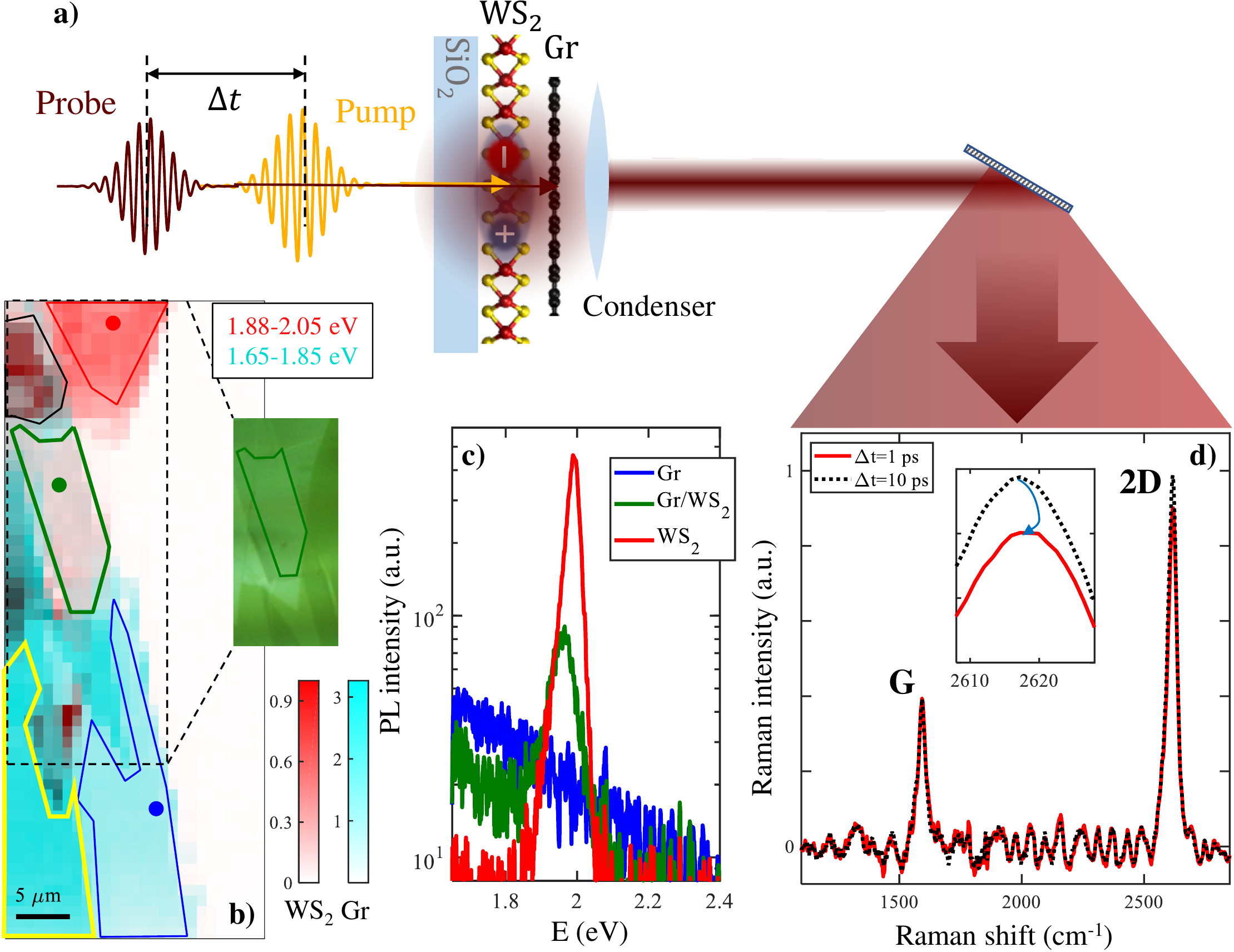} 
	\caption{\textbf{Optical Characterization of a WS$_2$-Gr heterostructure.} 
		a) Time delayed pump and probe beams focused onto a diffraction-limited WS$_2$-Gr spot. b) PL map generated by probe only (1.58~eV) after integrating the emission spectra in the ranges 1.88-2.05 eV (WS\textsubscript{2}, two-photon PL) and 1.65-1.85 eV (Gr, hot luminescence) and the corresponding optical image (right side).
			Contour lines: bare Gr (blue), few layer Gr (yellow), WS\textsubscript{2} (red), coupled (green) and uncoupled (black) WS$_2$-Gr regions.
			 c) PL spectra for the spots indicated in the map with corresponding colors.
			d) Time-resolved, probe generated Raman spectra of Gr modes in the WS$_2$-Gr heterostructure for two time delays, corrected from hot PL background (Materials and Methods).
		  } 	\label{fig_exp}
\end{figure}

A key technique for characterizing vdW heterostructures is Raman spectroscopy. The Raman spectrum of Gr is dominated by two features, known as the G and 2D modes, near 1580~cm$^{-1}$ and 2600~cm$^{-1}$, respectively. The G- and 2D-mode processes arise from non-resonant inelastic scattering with one zone-center longitudinal and transverse optical phonon and from a resonant process involving  a pair of near zone-edge transverse optical phonons, respectively \cite{ferrari_basko_2013,malard2009raman,Chen2011,Basko2008}. Both modes are sensitive to external perturbations and in particular to doping \cite{pisana,Pinczuk2007,Froehlicher2015} and temperature rises \cite{Bonini2007,freitag_2010,Berciaud2010}. For instance, the integrated intensity of the 2D-mode feature ($I_{\rm 2D}$) is routinely used to determine the Fermi level in  Gr \cite{BaskoA2D,Chen2011,Froehlicher2015} as well as of the defect density \cite{Venezuela2011} and the electronic temperature \cite{Giegold2020,Berciaud2010} as summarised in Eq. \eqref{eq:basko} below. Building upon this sensitivity, quantitative Raman-based methods have been developed to characterize Gr-based systems. Recently, a combination of photoluminescence (PL) spectroscopy and steady state spontaneous Raman spectroscopy have revealed the interplay between slow, extrinsic photoinduced net charge transfer and picosecond energy transfer, pointing towards the dominant role of energy transfer in TMD-Gr heterostructures \cite{Froehlicher2018}, without however determining the associated temporal dynamics.

Extending such sensing capabilities to the out-of-equilibrium regime, would offer a unique way to address the ultrafast response of Gr following photo-excitation of the TMD above its optical bandgap. Indeed, both charge and energy transfer  processes may in principle generate  hot carriers in Gr, leading to an increase in its electronic temperature $T_{\rm e}$, which can be quantified by monitoring hot PL and out of equilibrium Raman scattering. The former directly yields $T_{\rm e}$ \cite{Heinz2010}, while the latter is sensitive to both transient Fermi level shifts and temperature rises \cite{Ferrante2018,yan2009}.  
Assuming that photoexcitation of the TMD-Gr heterostructure predominantly yields a net charge flow (i.e., charge transfer) across the coupled layers, a long-lived doped state is expected to develop in Gr after electron-phonon relaxation, affecting the lineshape and the intensity of the G-mode and 2D-mode features. In contrast, in an energy transfer-dominated regime, Gr would only exhibit a transient increase of $T_{\rm e}$, which dynamics is governed by the interplay between energy transfer and phonon relaxation of the hot electron-hole (e-h) pairs injected in Gr occurred in the sub-picosecond timescale \cite{Butscher2007,Lazzeri2006,Tielrooij2013,Heinz2010}. This scenario would result in a sub-picosecond decrease of $I_{\rm 2D}$ combined with a broadening of the G-mode feature. Charge  and energy transfer can thus be distinguished based on the transient dynamics of the Raman features of Gr.

Here, we provide direct mechanistic and temporal insights into interlayer coupling in monolayer tungsten disulfide (WS$_2$)-Gr heterostructures by exploiting recent developments in ultrafast Raman spectroscopy \cite{Ferrante2018,virga_coherent_2019,Giegold2020,yan2009}.
Specifically, extracting the temporal profile of $T_{\rm e}$ in Gr following optical excitation of the WS$_2$ monolayer slightly above its optical bandgap, we demonstrate highly efficient energy transfer from TMD to Gr in a 4 ps timescale, with no evidence for a net charge transfer.

\section*{Results}

Our model system to track ultrafast interlayer transfers is a vdW heterostructure made from a Gr monolayer stacked onto  WS$_2$ monolayer supported by a silica substrate. In our experiments, the sample is \textit{probed} using  1 ps pulses at 1.58 eV, i.e. around 400 meV below the WS$_2$ optical bandgap. The role of the probe pulse is twofold: \textit{i)} it induces a combination of linear and non-linear PL which we use as an imaging tool for preliminary mapping the heterostructure. \textit{ii)}
it serves as a time-delayed Raman-inducing beam following photoexcitation by 1 ps pulses at 2.07 eV, slightly above the WS$_2$ optical bandgap. The time duration of the probe pulse is imposed by the necessary trade-off between a pulse duration ($\delta t$) as short as the exciton lifetime in the WS$_2$ and a narrow excitation bandwidth ($\delta \nu$) required to yield a sufficient spectral resolution (these two quantities are fundamentally constrained by the Fourier relation: $\delta \nu \delta t \geq 14.7~\mathrm {cm^{-1}} \mathrm{ps}$). The experiment concept is shown in Fig.~\ref{fig_exp}a,d and further detailed in the Method section.

\subsection*{Photoluminescence} 

PL spectroscopy is one of the main ways to characterize the TMD-Gr coupling. Bare TMD monolayers exhibit indeed strong PL: the radiative recombination upon optical excitation is favored by the presence of a direct bandgap in absence of competitive non-radiative decay channels \cite{mak2010,splendiani}. Conversely, Gr does not exhibit significant PL \cite{liu2016van} as efficient non-radiative decay channels (electron-phonon coupling) are active at any excitation wavelength owing to its gapless band structure  \cite{Gokus2009}. Accordingly, massive quenching  of the TMD PL indicates  efficient near-field coupling to Gr \cite{He2014,massicotte,Froehlicher2018,pierucci,Yuan2018} (see also Fig.~\figScwPL). 

A different scenario arises when photoexciting Gr with ultrashort laser pulses. Under  large photon fluxes, the non-radiative recombination channel is significantly depleted and hot-PL (albeit weaker than the TMD PL) is observed from Gr \cite{Heinz2010,Stohr2010}. Here, we take advantage of such transient regime to introduce an imaging protocol, based on the PL detection upon pulsed laser excitation, for simultaneously mapping Gr and WS$_2$. Specifically, photoexciting the WS$_2$-Gr heterostructure below the WS\textsubscript{2} optical bandgap using our probe pulse,  the WS\textsubscript{2} PL stems from a lower-efficiency two-photon absorption process(see Fig.~\figSPLtwoone), whose intensity is comparable (as shown in Fig.~\ref{fig_exp}c) with the Gr hot-PL and is emitted in a different spectral region.

By recording a map of the integrated PL intensity in distinct spectral regions selected to maximize the contrast (see Fig.~\ref{fig_exp}b), we are able to isolate the regions with Gr and WS\textsubscript{2} only, as well as the heterostructure with strongly and weakly coupled areas. Such simultaneous mapping, not achievable with 
continuous wave (cw) excitation, is a powerful method for imaging Gr-based heterostructures. The green contour of Fig.~\ref{fig_exp}b highlights the region of coupled WS$_2$-Gr, characterized by PL  quenching relative to bare WS\textsubscript{2}  and by the presence of  hot-PL from the coupled Gr monolayer (Fig.~\ref{fig_exp}c). The Raman measurements presented hereafter are performed in this coupled region of the heterostructure.

\subsection*{Time-resolved Raman scattering}

In our experiments the Raman signal induced by probe pulses is used to monitor Gr response following optical excitation by the higher energy pump photons absorbed by the WS$_2$ monolayer. Probe pulses also lead to an out-equilibrium  increase of the electronic temperature $T_{\rm e}^{\rm pr}\sim 1550~\rm K$ in Gr (Materials and Methods), due to the e-h pairs injected upon absorption ($A_{\rm Gr}\sim 2.3\%$). In contrast, the pump pulses  are mainly absorbed by the WS$_2$ layer (optical absorptance $A_{\rm X}\sim 5.4 \%$ at 2.07 eV in the same preparation condition \cite{Li2014,doi:10.1021/nl503799t}), generating excitons in WS$_2$. However, a fraction of the pump pulse is absorbed also by the Gr layer ($A=(1-A_{\rm X})A_{\rm Gr}\sim 2.2\%$), and induces a photo-induced electronic heating of Gr \cite{Heinz2010} (for 70 $\mu$W, $T_{\rm e}<$1400 K, see Fig. \figSfitHOTPL, implying an e-h pair density $n_{\rm eh}<4.5 \times 10^{11}$ cm$^{-2}$, see Fig. \figSTeneh). Importantly, probe pulses have a negligible interaction with WS$_2$, as testified by the weak two-photon fluorescence (see Fig. \figSPLtwoone). 

\begin{figure}
	\centering
	\includegraphics[width=0.6\linewidth]{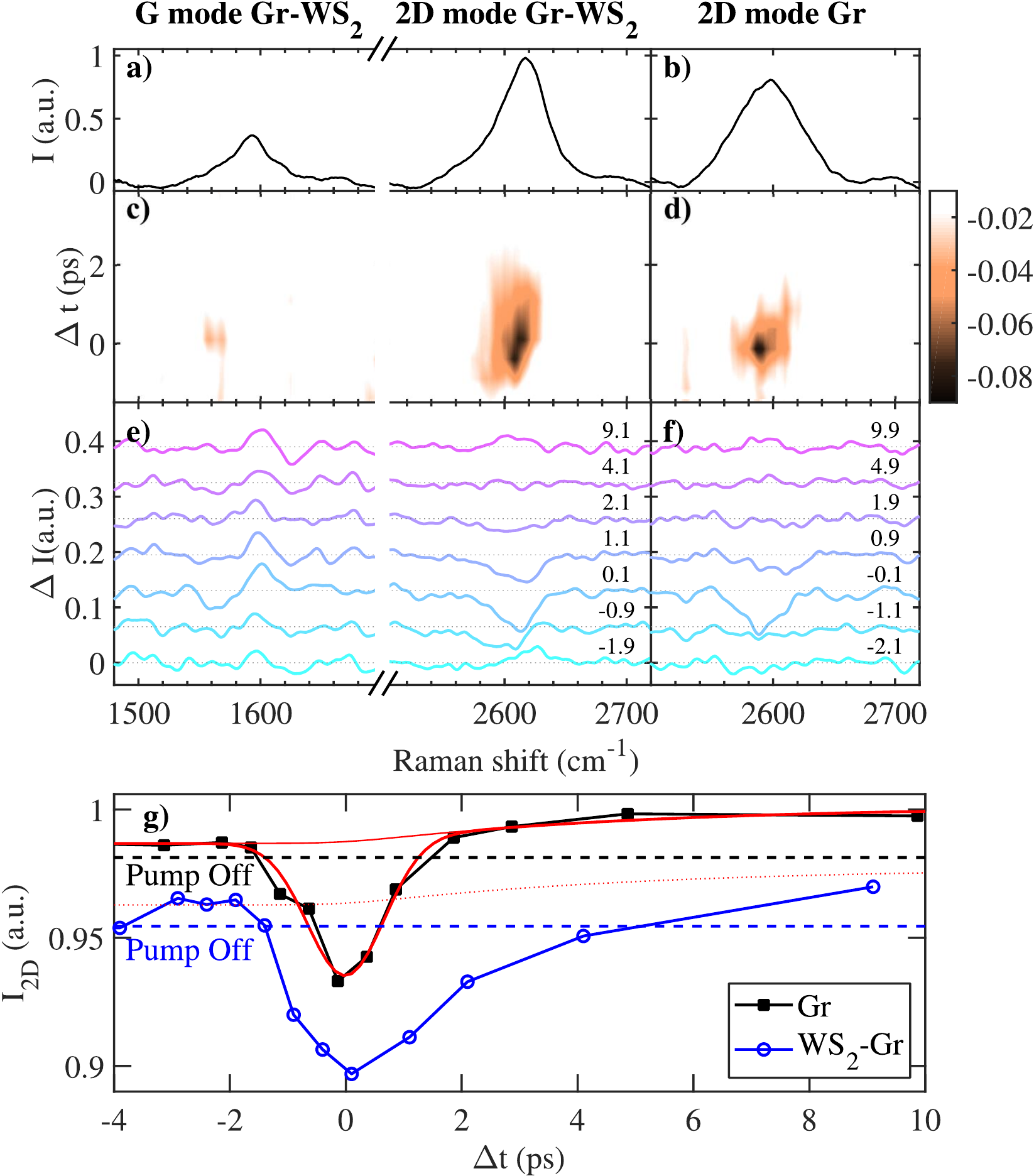} 
	\caption{\textbf{Time-resolved Raman spectra}. (a-b) Pump-off Raman spectra of the G and 2D modes for WS\textsubscript{2}-Gr and bare graphene (Gr). Transient differential Raman spectra $\Delta I(\Delta t,E)=I(\Delta t,E)-I(-30 ps,E)$: colormaps (c-d) and vertically offset slices for selected time delays in ps (e-f). The 2D-mode intensity $I_{\rm 2D}$, as opposed to $I_{\rm G}$, decreases around zero delay. This effect is observed to a lesser extent in bare Gr. (g) While $I_{\rm 2D}$ drop -due to electronic heating- recovers in bare Gr within a timescale comparable with the pump-probe temporal overlap (black symbols and guideline), it takes longer in WS$_2$-Gr (blue symbols and guideline). The dashed blue (black) lines show $I_{\rm 2D}$ without the pump beam in WS$_2$-Gr (bare Gr), indicating the presence of a small photo-induced phonon heating. The bare Gr profile  has been modeled (red thick line) with a picosecond drop (fast electronic term broadened by instrumental resolution) and its convolution with an exponential term (transient phononic contribution, red thin line): 
$f(\Delta t)=C+A\exp\left[ -\Delta t^2 / \left(2 \sigma^2\right) \right] + \Delta I_{\rm 2D}^{\rm ph}(\Delta t)$, where $\Delta I_{\rm 2D}^{\rm ph}(\Delta t)=B \exp\left( -\Delta t^2 / 2 \sigma^2 \right)\circledast \left[\theta(\Delta t) \left(1-e^{-\left(\Delta t\right) / \tau} \right)\right]$. 
$A$, $B$, $C$ are fitting parameters. $\sigma=0.66$ ps corresponds to the autocorrelation of the 1~ps FWHM pump and probe pulses, $\tau=5$ ps from \cite{Bonini2007}. $\Delta I_{\rm 2D}^{\rm ph}$ is also reported with a vertical offset to emphasize its role in the WS$_2$-Gr case (dotted red line). The accuracy in the $I_{\rm 2D}$ measurement is $\sim 1\%$.} 	\label{fig_spectra}
\end{figure}

Figure \ref{fig_spectra} shows time-resolved Raman spectra of the WS$_2$-Gr heterostructure compared with reference measurements performed on the nearby bare Gr region.
Our measurements without pump pulse reveal  a slightly lower value of $I_{\rm 2D}$ in WS$_2$-Gr, with respect to Gr, due to static charge transfer to Gr (dashed line of Fig.~\ref{fig_spectra}g), in agreement with cw experiments (Fig.~\figScwRaman d and \cite{Froehlicher2018,berciaud_2020}).
As shown in Fig.~\ref{fig_exp}d and \ref{fig_spectra}g, the integrated area of the 2D-mode feature  depends on the time delay. 
This effect appears prominently in Fig.~\ref{fig_spectra}c,d, where the differential 2D mode spectra (upon subtraction of the Raman spectra recorded at a $\Delta t = -30 ~\mathrm{ps}$)  exhibit a clear intensity depletion.
Moreover, Fig.~\ref{fig_spectra}g indicates the presence of a small pump-induced heat transfer from the electronic to the phononic degree of freedom, rising $I_{\rm 2D}$ above the pump off value (dashed lines of Fig.~\ref{fig_spectra}g) at large (due to transient D phonon heating) and at negative (increased lattice temperature) $\Delta t$.  In contrast, the differential spectra of the G-mode feature (Figs. \ref{fig_spectra}c, \figSGgr) do not exhibit any sizeable transient intensity change. Remarkably, while the transient decrease of $I_{\rm 2D}$ sets in and decays almost entirely within our temporal resolution in the Gr reference, the same effect decays on a longer timescale of $\sim 4 ~\rm{ps}$ in WS$_2$-Gr.

\section*{Discussion}
The PL quenching of WS\textsubscript{2} may stem either from charge transfer or from energy transfer.
Here, we numerically simulate the Raman experimental results considering only energy transfer from WS\textsubscript{2} to Gr with near-unity efficiency.

\begin{figure}
	\centering
	\includegraphics[width=12cm]{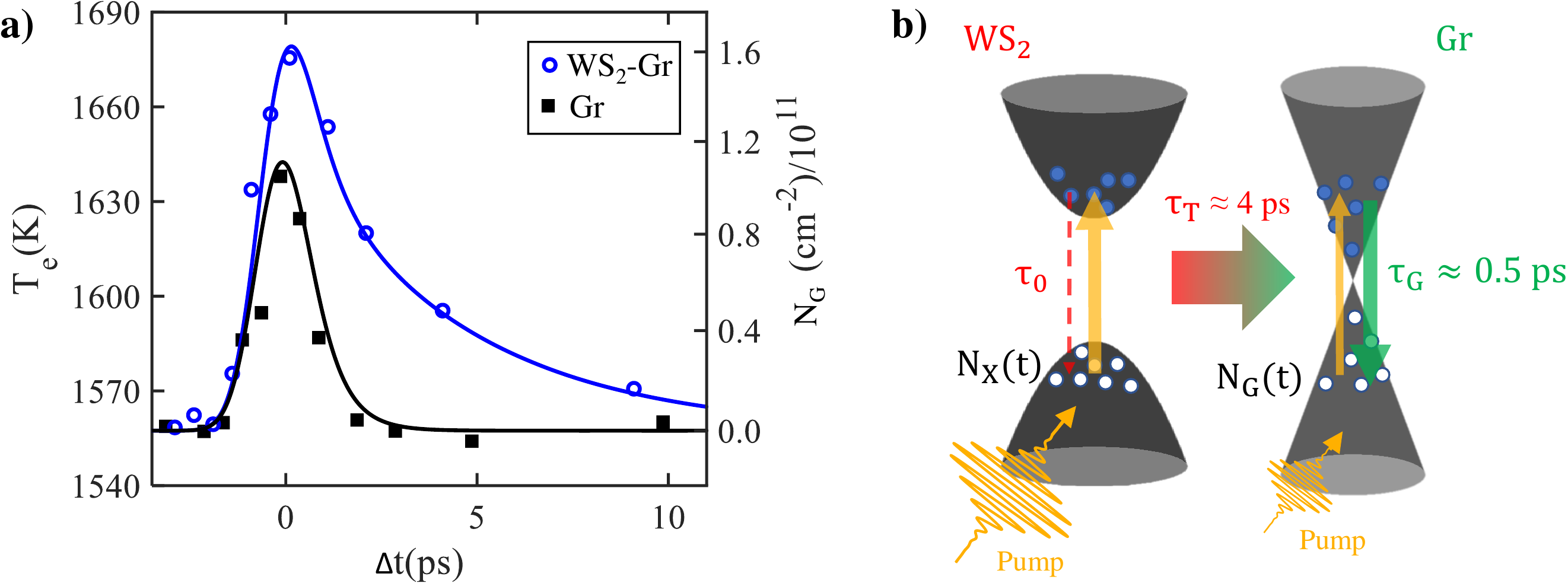} 
	\caption{\textbf{Modelling energy transfer in a WS\textsubscript{2}-Gr heterostructure.} a) $T_{\rm e}$ and e-h pair density at different time delays extracted from the dynamics of $I_{\rm 2D}$ for  WS\textsubscript{2}-Gr (open symbols) and bare Gr (full symbols) are compared with the simulated profiles (solid lines). The uncertainty on the $T_{\rm e}$ is 20 K.  
		b) Sketch of the kinetic model used in the simulation. The pump pulse can: (i) generate an exciton population in  WS\textsubscript{2} (tick orange arrow), (ii) populate the electronic states of Gr with e-h pairs (thin orange arrow). The e-h pairs in Gr decay with a time scale $\tau_{\rm G}$. In contrast, the excitons in bare WS\textsubscript{2} have a long lifetime $\tau_0$. Exciton decay is strongly accelerated in WS\textsubscript{2}-Gr due to energy transfer to Gr with a characteristic time $\tau_{\rm T}$.}
	\label{fig_model}
\end{figure} 

We first focus on the time-resolved Raman response of bare Gr. In this case, the fast drop and recovery of $I_{\rm 2D}$ can be ascribed to the e-h pair  density instantaneously photo-generated upon pump absorption and relaxing on an ultrafast (resolution limited) timescale, followed by a slower D phonon heating. 
In order to isolate this latter, the Gr experimental profiles have been preliminary fitted as the sum of a Gaussian dip and an exponential saturation term (see Fig. \ref{fig_spectra}), which takes into account for the electronic and subsequent lattice heating, respectively.
This latter ($\Delta I_{\rm 2D}^{\rm ph}$), expected to be weakly sensitive to the WS\textsubscript{2} coupling, is reported in Fig.~\ref{fig_spectra}g.
The genuine modification of the 2D area induced by the electronic heating ($I_{\rm 2D}^{\rm e}$) is hence obtained for both Gr and WS$_2$-Gr by subctracting the phononic term, as $I_{\rm 2D}^{\rm e}=I_{\rm 2D}-\Delta I_{\rm 2D}^{\rm ph}$. 

In order to quantitatively evaluate the e-h pair density injected by the probe pulse and the corresponding $T_e$ increase, the relation between the Gr electronic properties and the $I_{\rm 2D}$ can be exploited \cite{BaskoA2D}:
\begin{equation}
\begin{split}
I_{\rm 2D}^{\rm e}&\propto\left(\frac{\gamma_{\rm K}(E)}{\gamma(E,T_{\rm e})}\right)^2=\\
&=\left(\frac{\gamma_{\rm K}(E)}{\gamma_{\rm K}(E)+\gamma_\Gamma(E)+\gamma_{\rm def}(E)+\gamma_{\rm ee}(E,T_{\rm e})}\right)^2
\label{eq:basko}
\end{split}
\end{equation}
where $E$ denotes the energy of the electronic state and 2$\gamma$ is the scattering rate of electrons and holes, equal to the sum of   $\gamma_{\rm K}$, $\gamma_\Gamma$, $\gamma_{\rm def}$ and $\gamma_{\rm ee}$, which are the scattering rate of electrons with 2D-mode and G-mode phonons near the K and $\Gamma$ points of the Brillouin zone, defects and other electrons, respectively. Noteworthy, $\gamma_{\rm ee}$ is the only term that depends on $T_{\rm e}$ \cite{pisana,Pinczuk2007,Neumann2015,Bonini2007,Venezuela2011,BaskoA2D}. Specifically \cite{Schutt2011} 
\begin{equation}
\gamma_{\rm ee}(E,T_{\rm e})=\alpha T_{\rm e}
\label{eq:gam_te}
\end{equation}
with $\alpha =0.51$~cm$^{-1}$K$^{-1}$ for $E>100$~meV \cite{Ferrante2018}.

From the full width at half maximum (FWHM) of 2D-mode ($\Gamma_{\rm 2D}$)   measured without pump pulse, we can estimate $\gamma \approx 195~\rm{meV}$  for  WS\textsubscript{2}-Gr at $T_{\rm e}=T_{\rm e}^{\rm pr}$ \cite{Basko2008}. In the presence of pump pulse, an increase of $T_{\rm e}$ is expected. However, for negative or large pump-probe delays ($>4$~ps) the electronic heating of Gr induced by the direct interaction with pump pulse is not observed within the probe pulse duration due to the ps recombination of the e-h pairs. 
The electronic heating effect can only be observed around the time-overlap of pump and probe pulse, generating a drop in $I_{\rm 2D}^{\rm e}$ (see Fig.~\ref{fig_spectra}g). 
Taking into account this experimental evidence and Eq. \eqref{eq:basko}, the temporal profile of $T_{\rm e}$ in WS$_2$-Gr and in bare Gr can be extracted (Fig.~\ref{fig_model}a, Materials and Methods). 

As shown in Fig.~\ref{fig_model}a, the additional energy flux injected through energy transfer  from the WS\textsubscript{2} layer to Gr leads to a different $T_{\rm e}$ profile as compared to reference measurements in bare Gr, taking into account that the energy associated with a WS$_2$ exciton transfer is $\sim2$ eV and using the Fermi distribution with the density of states in Gr, we can derive the relation between the additional e-h pair density induced by the pump pulse and the relative increase of $T_{\rm e}$ above $T_{\rm e}^{\rm pr}$ (right axis of Fig.~\ref{fig_model}a, Materials and Methods).

At this point, we model the dynamics of e-h pairs in Gr taking into account the transfer pathways reported in Fig.~\ref{fig_model}b.  The pump pulse creates a population of e-h pairs in the WS$_2$ and Gr monolayers proportional to their absorptances.
The WS\textsubscript{2} layer with Gr on top exhibits a fast decay channel ($\tau_{\rm T}$), which inhibits its much longer decay $\tau_{0} \sim 100~\rm{ps}-1~\rm{ns}$ \cite{Froehlicher2018,Yuan2018,He2014,Lorchat2018} and exciton-exciton annihilation \cite{berciaud_2020}. Due to the suppression of 
this slow linear recombination in bare WS$_2$, the temporal evolution of the exciton density in WS\textsubscript{2}, $N_{\rm X}(t)$, can be described by the following differential equation:
\begin{equation}
\frac{dN_{\rm X}}{dt}=-\frac{N_{\rm X}}{\tau_{\rm T}}+A_{\rm X}I_{\rm PUP}(t)
\label{eq_diffTMD}
\end{equation} 
where $I_{\rm PUP}(t)$ is the temporal profile of the pump pulse, assumed Gaussian with a FWHM of 1 ps.

On the other hand, Gr exhibits an intrinsic decay time for the thermalisation of the electronic system with the phononic bath and substrate ($\tau_{\rm G}$).
Interlayer energy transfer induces e-h pair injection in Gr ($N_{\rm G}(t)$), leading to a rise in $T_{\rm e}$.
Its temporal evolution can be expressed as:
\begin{equation}
\frac{dN_{\rm G}}{dt}=-\frac{N_{\rm G}}{\tau_{\rm G}}+ \frac{N_{\rm X}}{\tau_{\rm T}}+A I_{\rm PUP}(t)
\label{eq_diffG}
\end{equation}

where the coupling term $\frac{N_{\rm X}}{\tau_{\rm T}}$ obviously vanishes in the bare Gr case. Performing a best fit of the experimental data reported in Fig.~\ref{fig_model}a with $N_{\rm G} (t)$ obtained from Eq. \eqref{eq_diffTMD} and Eq. \eqref{eq_diffG}, convoluted with the 1 ps probe pulse temporal profile, we obtain the black and blue solid lines for bare Gr and WS$_2$-Gr, respectively, with characteristic times $\tau_{\rm G}=0.6$~ps, $\tau_{\rm T}=4.3$~ps.

The decay time $\tau_{\rm G}$ in bare Gr is close to the temporal resolution of our setup and is consistent with previous reports on Gr \cite{shiwei2012,PhysRevB.83.121404,PhysRevLett.95.187403}. Most importantly, we are able to determine the relevant energy transfer timescale from WS$_2$ to Gr at room temperature. Remarkably, $\tau_{\rm T} \approx 4~\rm {ps}$  is longer than $\tau_{\rm G}$, indicating that energy transfer-mediated e-h pair injection in Gr (the term $\frac{N_{\rm X}}{\tau_{\rm T}}$ in Eq. \eqref{eq_diffG}) is responsible for the longer decay of $N_{\rm G}(t)$ observed in WS$_2$-Gr.  

Since we are investigating a  system separated by a sub-nm van der Waals gap, F\"orster and Dexter energy transfer are both likely to occur with comparable efficiencies. The $\tau_{\rm T}$ value that we directly determine here, however, provides a possible rationale to previous observations of PL quenching \cite{Froehlicher2018} and TA spectroscopy~\cite{He2014,Yuan2018,Chen2019} and agrees with theoretical calculations based on a dominant F\"orster-type energy transfer mechanism \cite{Selig2019}.

Noteworthy, the WS\textsubscript{2} PL quenching and the $I_{\rm 2D}$ drop \cite{BaskoA2D} may be compatible with a charge transfer mechanism involving a net transient flux of electrons (or holes) to Gr.
Critically, in this scenario, the doping generated in Gr upon charge transfer would be expected to relax on a time scale much longer than the picosecond dynamics reported in Fig.~\ref{fig_spectra}g \cite{Froehlicher2018,Ahmed2020}. Moreover, the measured temporal profile of the G band FWHM  is also not compatible with a charge transfer-dominated scenario, as detailed in the Supplemental Material. 

The energy transfer nature of the process established here suggests that the photogating effects \cite{Froehlicher2018,Lin2019} and associated photodetection capabilities demonstrated in TMD-Gr-based optoelectronic devices  \cite{Zhang2014, Ahmed2020} result from an additional conversion mechanism into a much slower, less efficient, net charge transfer.
Moreover, this evidence confirms the dominant contribution of energy transfer, measured with Raman spectroscopy  \cite{Froehlicher2018} and at the same time the ps simultaneous transfer of e-h of the exciton  provides an explanation for the low internal quantum efficiency of photocurrent generation in photodetectors made from TMD monolayers  \cite{massicotte}.

\section*{Conclusion}

In conclusion, we have measured the time-resolved Raman spectrum of graphene in a WS\textsubscript{2}-Gr heterostructure  upon resonant photoexcitation of band-edge WS\textsubscript{2} excitons. 
By comparing the Raman response of bare and WS\textsubscript{2}-coupled Gr, we have unveiled a slower relaxation of the 2D-mode integrated intensity occurring on a 4~ps  timescale in WS\textsubscript{2}-Gr. Our experimental data are rationalized by a kinetic model based solely on energy transfer. Although photocurrent generation occurs with relatively high efficiency in vertically biased TMD-Gr heterojunctions \cite{massicotte,Arp2019} made from few-layer TMD films, our results strongly suggest that net photo-induced charge transfer has negligible efficiency in ultimately thin monolayer TMD-Gr heterostructures, explaining the low efficiency of associated photo-devices.
Instead, we demonstrate that a TMD monolayer can boost carriers injection in Gr through picosecond energy transfer, with a room-temperature transfer yield approaching unity considering the longer excitonic lifetimes in bare TMD monolayers. Beside showing the possibility to exploit photothermionic effects for efficient hot carrier injection from Gr to TMDs under sub-bandgap excitation \cite{massicotte2016,Chen2019}, our results provide an essential step for the microscopic understanding of ultrafast interlayer coupling in TMD-Gr heterostructures, whose relevance for optoelectronics is thus consolidated. They also indicate as a key technological challenge the efficient funneling of hot carriers generated in Gr before they release their energy into heat, either through Gr phonon emission or through coupling with the substrate.

\section{Methods}

\subsection*{Sample preparation}
Our WS$_2$-graphene heterostructure was fabricated as described previously~\cite{ Froehlicher2018} using a dry, viscoelastic transfer technique~\cite{Castellanos2014}. In brief, a WS$_2$ monolayer and a graphene monolayer were mechanically exfoliated from bulk crystals and stacked onto a fused silica substrate using a home-built transfer station. The sample was characterized using cw photoluminescence and Raman spectroscopies in ambient conditions (see Figs. \figScwPL~and \figScwRaman, respectively) before performing time-resolved studies. 

\subsection*{Photoluminescence measurements}
The experimental setup is the same as the time-resolved Raman experiment, but in this case only the probe pulse is sent on the sample. The PL spectra reported in Fig.~1b-c are collected in 2 sec per pixel and the sample scan is performed with two mechanical stages. 
This experimental scheme allows to perform a simultaneous imaging of the two monolayers. Moreover, using the non-linear two-photon PL for WS\textsubscript{2}, the spatial resolution is higher as compared to measurements in the cw regime performed using the same optical elements.
\\
Graphene having no band gap\cite{RevModPhys.81.109} is not expected to have any radiative decay channel for charge carriers. However, under high excitation densities it  emits light~\cite{Heinz2010} over a wide spectral range owing to the inhibition of a non-radiative recombination channel. Indeed, in an out-of-equilibrium configuration, the large density of charge carriers in the conduction band cannot fully relax down to the Fermi level via electron-phonon decay pathways. Thus, the electronic sub-system also relaxes radiatively through hot PL, which is well described by Planck's law.
Hence, in order to extract the electronic temperature ($T_{e}$), we fit the hot PL spectra of graphene with Planck's law (see black lines in Fig.~\figSHOTPL):

\begin{equation}
I(\hbar \omega,T_{e})=\eta_{\rm em} \tau_{\rm em}E(\hbar \omega)  \frac{\hbar \omega^3}{2\pi^2c^2}\frac{1}{e^{\frac{\hbar \omega}{k T_{e}}}-1} 
\label{planck}
\end{equation}
where $\eta_{\rm em}$ is the emissivity of the blackbody, $\tau_{\rm em}$ is the emission time of the blackbody and $E(\hbar \omega)$ is the responsivity of the detection chain taking into account the efficiency of the CCD which depends on the photon energy. $\hbar \omega$ is the energy of the emitted photon, $T_{e}$ is the electronic temperature and $k$ is Boltzmann's constant.

In  Fig.~\figSHOTPL ~we show the hot PL spectra obtained performing a probe pulse only experiment on Gr (b) and on WS\textsubscript{2}-Gr (a) at different powers of the probe pulse. The peak located near 1.8 eV is present in both Gr and in WS$_2$-Gr and is assigned to the anti-Stokes Raman G mode, while the broad peak at around 2 eV is assigned to PL from the A-exciton of WS\textsubscript{2} excited by a two-photon absorption process.
\\
In Fig.~\figSfitHOTPL ~we show the temperature extracted from fit as a function of probe pulse powers for both bare Gr (diamond) and WS$_2$-Gr (square).  
From the same figure, we  estimate that the probe pulses (with power of 210~$\mu$W) used in our time-resolved Raman scattering experiment lead to an electronic temperature $T_{\rm e}^{\rm pr}\sim1550$~K. The temporally overlapped pump pulses induce an additional temperature increase, originated from both direct electronic heating of Gr and energy transfer from  WS\textsubscript{2}.

\subsection*{Time-resolved Raman experiments}
We use a two-module Toptica FemtoFiber Pro source as in Ref. \cite{virga_coherent_2019}. This setup is able to generate, at a repetition rate of 40~MHz,  
1~ps probe pulses at 1.58~eV (PRP) and a supercontinuum (SC) output between 0.89-1.38~eV. The SC spectral intensity can be tuned with a motorized Si-prism-pair compressor. A Periodically Poled Lithium Niobate (PPLN) crystal with a fan-out grating (a poling period changing along the transverse direction) is exploited to produce broadly tunable (from 1.97~eV to 2.30~eV) narrowband 1~ps pump pulses , with a power$< 10$~mW \cite{Moutzouris}. A dichroic mirror is used to combine the two beams, whose relative temporal delay is tuned with an motorized optical delay line (DL).  A long-working distance 20$\times$ objective (numerical aperture $\rm{NA}=0.4$) focuses the pulses onto a same, $\sim 2 \mu$m spot on the sample.
Before objective the probe and pump power are 210~$\mu$W and 70~$\mu$W, respectively.
The Raman emission on the Stokes side is collected by a condenser (numerical aperture 0.75) and the pump-probe pulses are filtered out using interference filters. The Raman spectrum is measured with a monochromator (Acton Spectra Pro 2500i) coupled to a 
cooled CCD array (Princeton Instruments Pixis 100). 

In Fig.~2 the Stokes Raman signals of Gr generated by PRP are reported  at several time delays first on Gr and then on WS\textsubscript{2}-Gr. For each delay we acquire Raman spectrum for 10 minutes and repeat the acquisition 6 times.
In \HeteroSpectrum ~we show a typical spectrum in WS\textsubscript{2}-Gr with and without pump.
Raman spectra are affected by an intense background due to the substrate.
In order to remove this background, we collect, with same acquisition time, the signal spectrum from the substrate and we subtract it (with a scaling of the intensity) from Raman signal. In \HeteroSpectrumone(a) we show two selected spectra (pump off and 0.1~ps) in which we have removed the substrate background. In order to better clean data we remove a baseline (red) and get the spectra as in  \HeteroSpectrumone(b). We do the same procedure on the spectra got on Gr alone.   
The two peaks at $\sim 1585$ cm$^{-1}$ and $\sim 2620$ cm$^{-1}$ 
are the G and 2D modes, respectively.

\subsection*{\boldmath$T_{\rm e}$ and e-h pair density calculation}
To extract the $T_{\rm e}$ from Eq. (1), the value of $\gamma(T_{\rm e}^{\rm pr})$ is required, that can be extracted from the full-width at half maximum of the 2D-mode feature, $\Gamma_{\rm 2D}$\cite{Basko2008} that writes:
\begin{equation}
\Gamma_{\rm 2D}=4\sqrt{2^{2/3}-1}\frac{1}{2}\frac{\partial \omega_{\rm 2D}}{\partial (h\nu_\mathrm{laser})} \gamma(T_{\rm e})
\label{eq:gamma2D}
\end{equation}

where $\omega_{\rm 2D}$ is the 2D-mode frequency, $[{\partial \omega_{\rm 2D}}/{\partial (h\nu_\mathrm{laser})}]/2 =\frac{1}{c\: h} v_{ph}/v_\mathrm{F}\sim~100$~cm$^{-1}$ eV$^{-1}$,
\cite{Berciaud2013,ferrari_basko_2013} i.e. the ratio between the phonon and Fermi velocity, defined as the slope of phononic (resp. electronic) dispersion at the phonon (resp. electron) momentum corresponding to a given excitation laser energy $h \nu_\mathrm{laser}$\cite{ferrari_basko_2013}. $c$ and $h$ indicate the speed of light and the Planck constant, respectively.
However, eq. \ref{eq:gamma2D} does not consider the bimodal spectral feature of the 2D mode, due to the contributions of inner and outer processes\cite{Maultzsch2004,doi:10.1063/1.4729407}. For this reason, we isolate each contribution, following the procedure reported in Ref.~\cite{Berciaud2013}. The fitted bandwidth of the 2D mode contributions, obtained in cw measurements at room temperature, is 19~cm$^{-1}$, which we use to estimate $\gamma(300~K) =940 \rm{cm}^{-1}$.
Considering Eq. (2), $\gamma(T_{\rm e})=1600$ cm$^{-1}$ at $T_{\rm e}^{\rm pr}=1550$~K, as estimated above. 
To evaluate $T_{\rm e}$ from $I_{\rm 2D}^{\rm e}$, eq. (2) is replaced in eq. (1), obtaining: 
\begin{equation}
I_{\rm 2D}^{\rm e}(T_{\rm e})\propto\left(\frac{\gamma_K(E)}{\gamma(T_{\rm e}^{\rm pr})+\alpha (T_{\rm e}-T_{\rm e}^{\rm pr})}\right)^2
\label{eq:basko_mod}
\end{equation}
The ratio between $I_{\rm 2D}^{\rm e}$ at different $T_{\rm e}$ is:
\begin{equation}
\frac{I_{\rm 2D}^{\rm e}(T_{\rm e})}{I_{\rm 2D}^{\rm e}(T_{\rm e}^{\rm pr})}=\left(\frac{\gamma(T_{\rm e}^{\rm pr})}{\gamma(T_{\rm e}^{\rm pr})+\alpha (T_{\rm e}-T_{\rm e}^{\rm pr})}\right)^2
\label{eq:basko_rat}
\end{equation}
$T_{\rm e}$ extracted from the previous equation is:
\begin{equation}
T_{\rm e}-T_{\rm e}^{\rm pr}=\frac{\gamma(T_{\rm e}^{\rm pr})}{\alpha}\left(\sqrt{\frac{I_{\rm 2D}^{\rm e}(T_{\rm e}^{\rm pr})}{I_{\rm 2D}^{\rm e}(T_{\rm e})}}-1\right)
\label{eq:T_e_ext}
\end{equation}
Summing up, the latter equation is obtained by the dependence of  $I_{\rm 2D}^{\rm e}$ on  $T_{\rm e}$ (eq. \ref{eq:basko_mod}). The ratio between  $I_{\rm 2D}^{\rm e}$ at different temperatures allows to remove the dependence on $\gamma_K(E)$ (eq. \ref{eq:basko_rat}) and $T_{\rm e}$ can be conveniently isolated (eq. \ref{eq:T_e_ext}), considering the $\gamma(T_{\rm e}^{\rm pr}) =1600$ cm$^{-1}$.
However, the model in Fig.~3a is designed to reproduce the e-h pairs dynamics. Consequently a conversion of $T_{\rm e}$ in e-h pairs ($n_{\rm eh}$) is required.
In our experimental configuration the pump photon energy ($E_{\rm PUP}$) matches the TMD bandgap ($\sim 2$eV), and therefore the energy of each e-h pair either generated from the pump photons or transferred from WS\textsubscript{2}.
Hence, the carrier density is related to the total energy density (per unit surface) associated to a given $T_{\rm e}$ as:

\begin{equation}
\begin{split}
n_{\rm eh}=\frac{\varepsilon_{\rm TOT}(T_{\rm e})}{E_{\rm PUP}}&=\frac{1}{E_{\rm PUP}}\int_{-\infty}^{\infty} g(E)f(E,E_{\rm F},T_{\rm e}) dE =\\
&=\frac{1}{E_{\rm PUP}}\int_{-\infty}^{\infty}\frac{2 |E|}{\pi \hbar^2 v_{\rm F}} \frac{1}{e^{(E-E_{\rm F})/(k T)}+1}dE
\label{eq:Etot}
\end{split}
\end{equation}

where $g(E)$ is the density of states in Gr, $f(E,E_{\rm F},T_{\rm e})$ is the Fermi-Dirac distribution, $E_{\rm F}$ and $v_{\rm F}$ are the Fermi Energy and Fermi velocity and $k$ is the Boltzmann constant. Spontaneous Raman measurements (Fig. \figScwRaman) indicates a relatively low doping below $6 \times 10^{11}$  cm$^{-2}$, i.e., a Fermi level $|E_{\rm F}| < 100$ meV at room temperature. Moreover, under our experimental condition, the probe pulse alone induces $T_{\rm e}=T_{\rm e}^{\rm pr}$ in graphene, further reducing the upper bound for the Fermi level down to 30 meV \cite{chae}. 
In Fig.~\figSTeneh, we report $n_{\rm eh}$ as a function of $T_{\rm e}$ for $E_{\rm F}=0$ meV and $E_{\rm F}=30$ meV. Such doping effect mainly results in a minor vertical shift, not relevant for the incremental carrier generation $N_G$ (additional to the probe induced $n_{\rm eh}$) associated to a temperature rise from $T_{\rm e}^{\rm pr}$ to the pump induced time dependent $T_{\rm e}$ (Fig.~\ref{fig_model}b).

\section*{Acknowledgements}
	TS acknowledge the support from the PRIN 2017 Project 201795SBA3 – HARVEST. 
	TS and GB are grateful for the `Progetti di Ricerca Medi 2019' grant by Sapienza~Universit\'a~di~Roma. 
	We acknowledge Francesco Mauri for fruitful discussions.
	We are grateful to the StNano clean room staff and to M. Romeo for technical support. 
	SB acknowledges support from Institut Universitaire de France (IUF) and benefited from from a ``Sapienza'' University visiting professorship grant 2017.  
	This project has received funding from Agence Nationale de la Recherche under grants 2D-POEM ANR-18-ERC1-0009, and ATOEMS ANR-20-CE24-0010. This work of the Interdisciplinary Thematic Institute QMat, as part of the ITI 2021 2028 program of the University of Strasbourg, CNRS and Inserm, was supported by IdEx Unistra (ANR 10 IDEX 0002), and by SFRI STRAT'US project (ANR 20 SFRI 0012) and EUR QMAT ANR-17-EURE-0024 under the framework of the French Investments for the Future Program.



\newpage
\renewcommand{\thefigure}{S\arabic{figure}}
\setcounter{figure}{0}    
\renewcommand{\theequation}{S\arabic{equation}}
\large{\textbf{\centerline{Supplementary Materials for:}\\
		Picosecond energy transfer in a transition metal dichalcogenide-graphene heterostructure revealed by transient Raman spectroscopy}}

\section*{Supplementary Text}
\subsection*{Fluence dependence of the G and 2D Raman peaks}

\normalsize{During electron-phonon thermalization, the energy initially injected into the electron bath is not evenly distributed among all phonon-modes.
Consequently, each phonon mode may have its own temperature dependence upon photoexcitation. The G-mode involves the $E_{2g}$ phonon located at the $\Gamma$ point, where the e-ph coupling is particularly efficient. Hence, the photo-excited electron bath results in a short thermalization time $<$1ps \cite{PhysRevLett.95.187403} and, over our time scale (a few ps), we must consider the electronic temperature $T_e$ and the phonon temperature at $\Gamma$ $T_{\mathrm{G}_{ph}}$ close to the equilibrium: $T_e \sim T_{\mathrm{G}_{ph}}$. Thus, the phonon population at $\Gamma$ progressively increases with fluence resulting in a rise of the G-mode intensity.

Differently, the Raman 2D-mode involves a pair of near zone edge (K point) optical phonons (D phonons) with opposite momenta \cite{ferrari_basko_2013}. These ``D phonons'' do not couple as efficiently as $\Gamma$ point (or G) phonons as it is not  possible to satisfy momentum and energy conservation with thermal electron-hole couples.
Indeed, the involved electrons and holes should have an energy corresponding to half the incoming photon energy (corresponding to an unrealistically high $T_e \sim $~10$^4$ K).
Hence, the D phonon population rise is uniquely achieved via the less efficient anharmonic coupling, implying a optical-phonon thermalization time of several ps \cite{PhysRevLett.95.187403,Bonini2007}, i.e. much longer than the timescale of our experiment. Consequently, the electronic temperature $T_e$ is much higher than the phonon temperature responsible for the 2D Raman mode $T_{\mathrm{D}_{ph}}$: $T_e \gg T_{\mathrm{D}_{ph}}$. 

Consequently, Eq.1 explains why the 2D mode intensity decreases with $T_e$ via the $\gamma_{ee}(T_e)$ dependence.

This is at ease with previous Raman measurement (see \ref{G_2D_review}) performed with a single 1-ps narrowband beam in single Layer-CVD graphene \cite{Ferrante2018}. The optical beam power increases the electronic temperature similarly to fig. 1b of ref \cite{Ferrante2018} in line with fig. \ref{figS_fitHOTPL} of the present manuscript. As shown in the Fig. \ref{G_2D_review}, the G mode intensity (normalized to the laser power) increases with the beam power/electronic temperature (in agreement with \cite{yan2009}). On the contrary, the 2D mode intensity (normalized to laser power) decreases with the beam power/electronic temperature in agreement with the $T_e \gg T_{\mathrm{D}_{ph}}$ condition.
\begin{figure}[htbp]
	\centering
	\begin{tabular}{cc}
		\includegraphics[width=7cm]{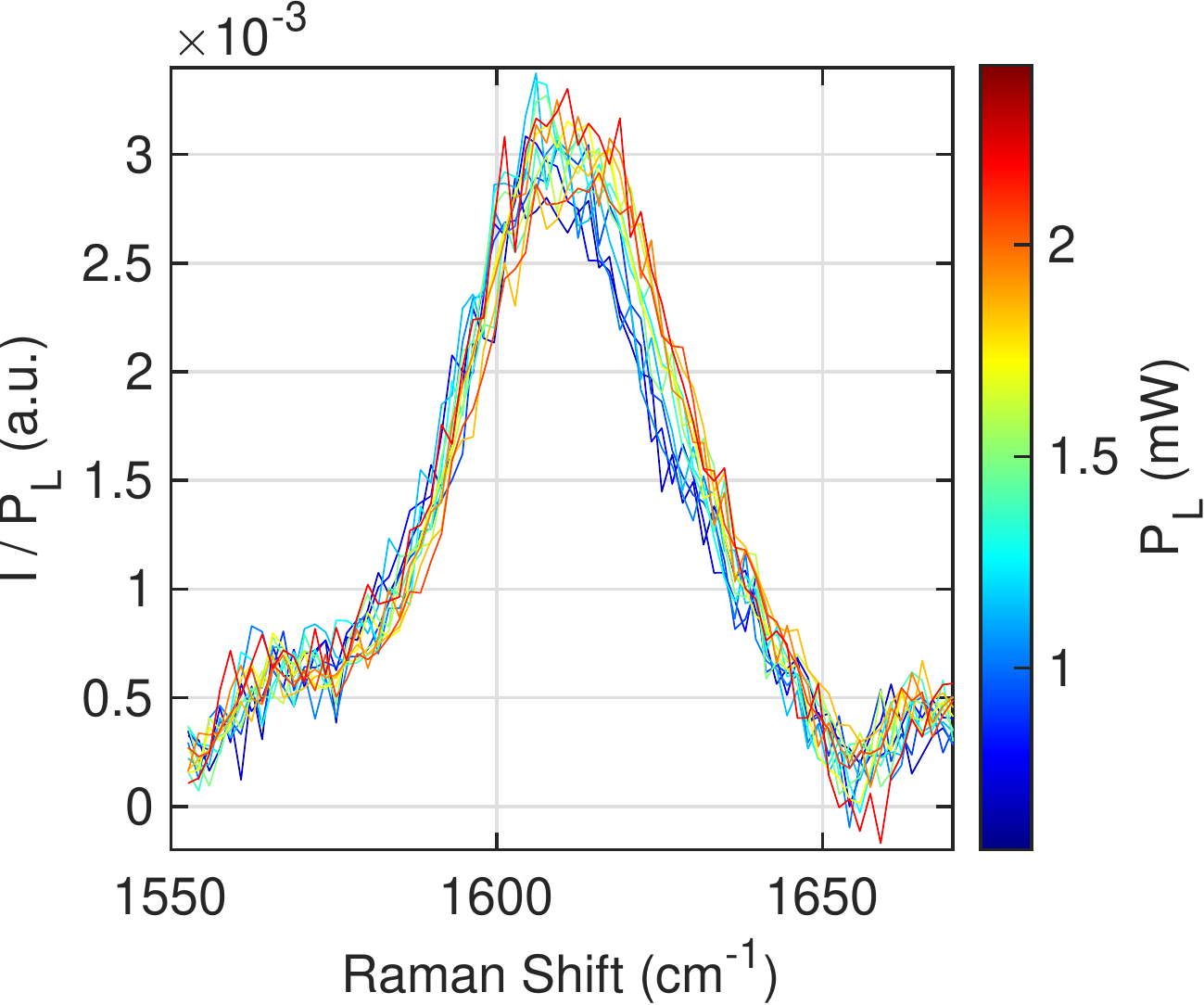}&
		\includegraphics[width=6.5cm,height=5.5cm]{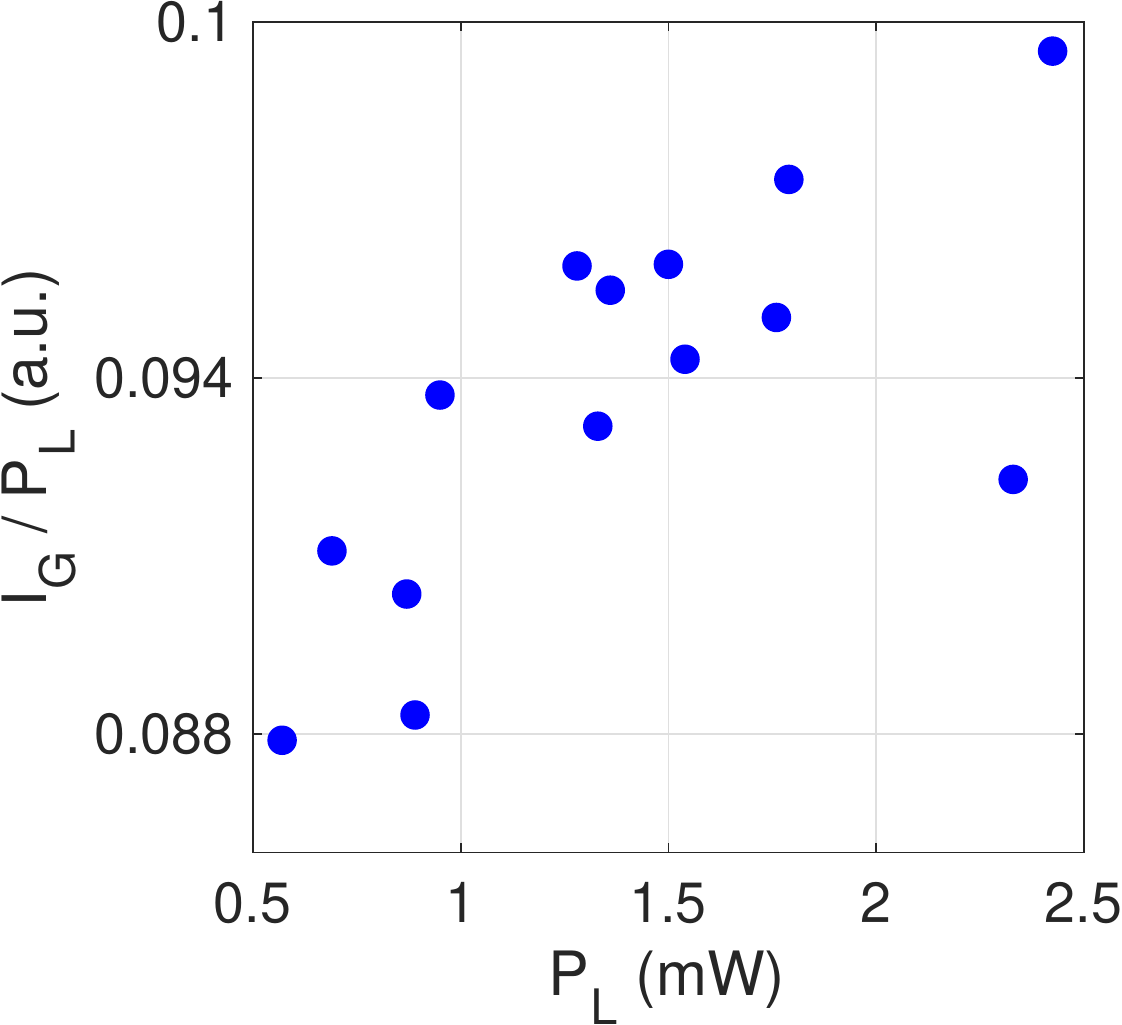}\\
		\textbf{a)}&
		\textbf{b)}\\
		\includegraphics[width=7cm]{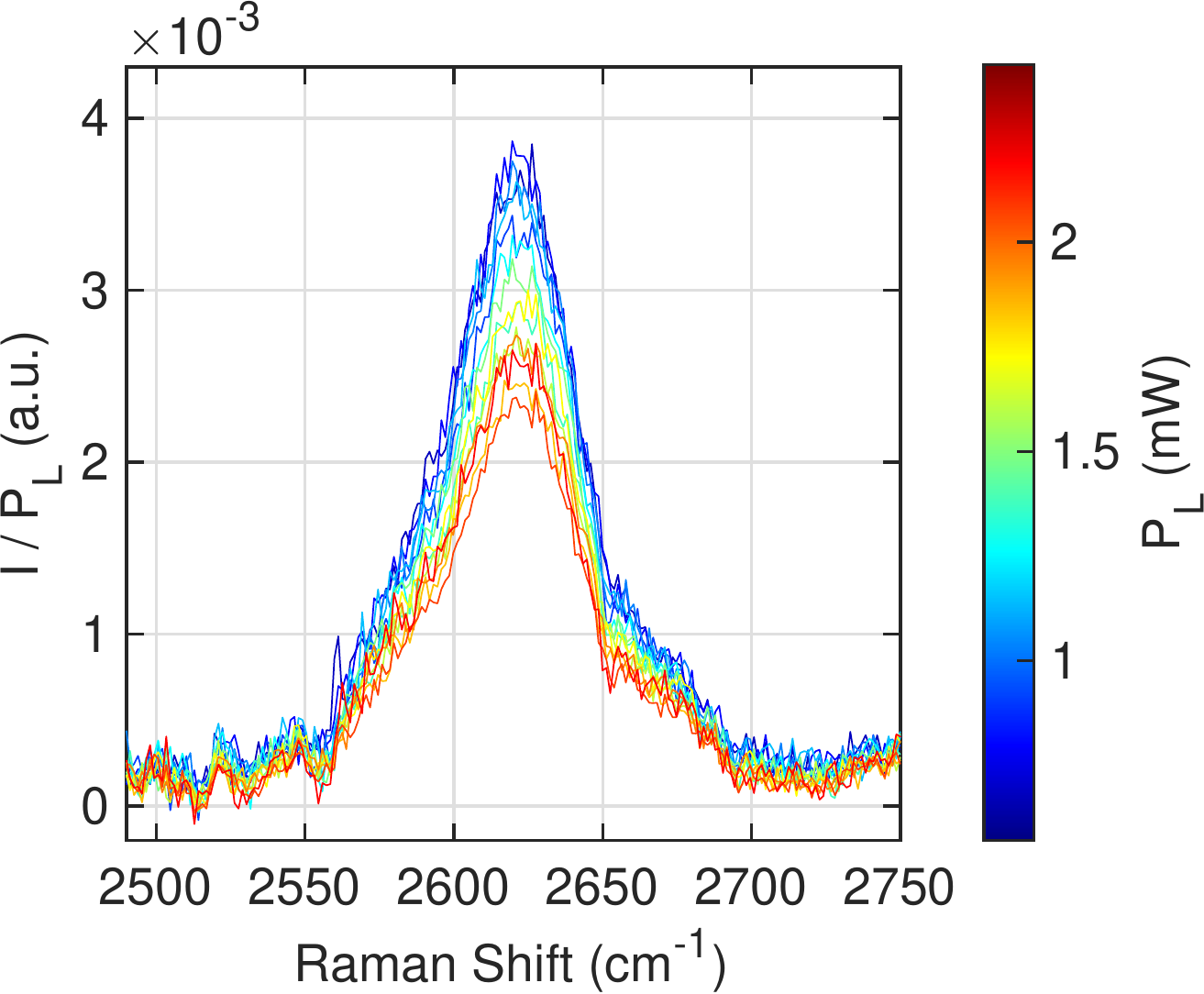}&
		\includegraphics[width=6.5cm,height=5.5cm]{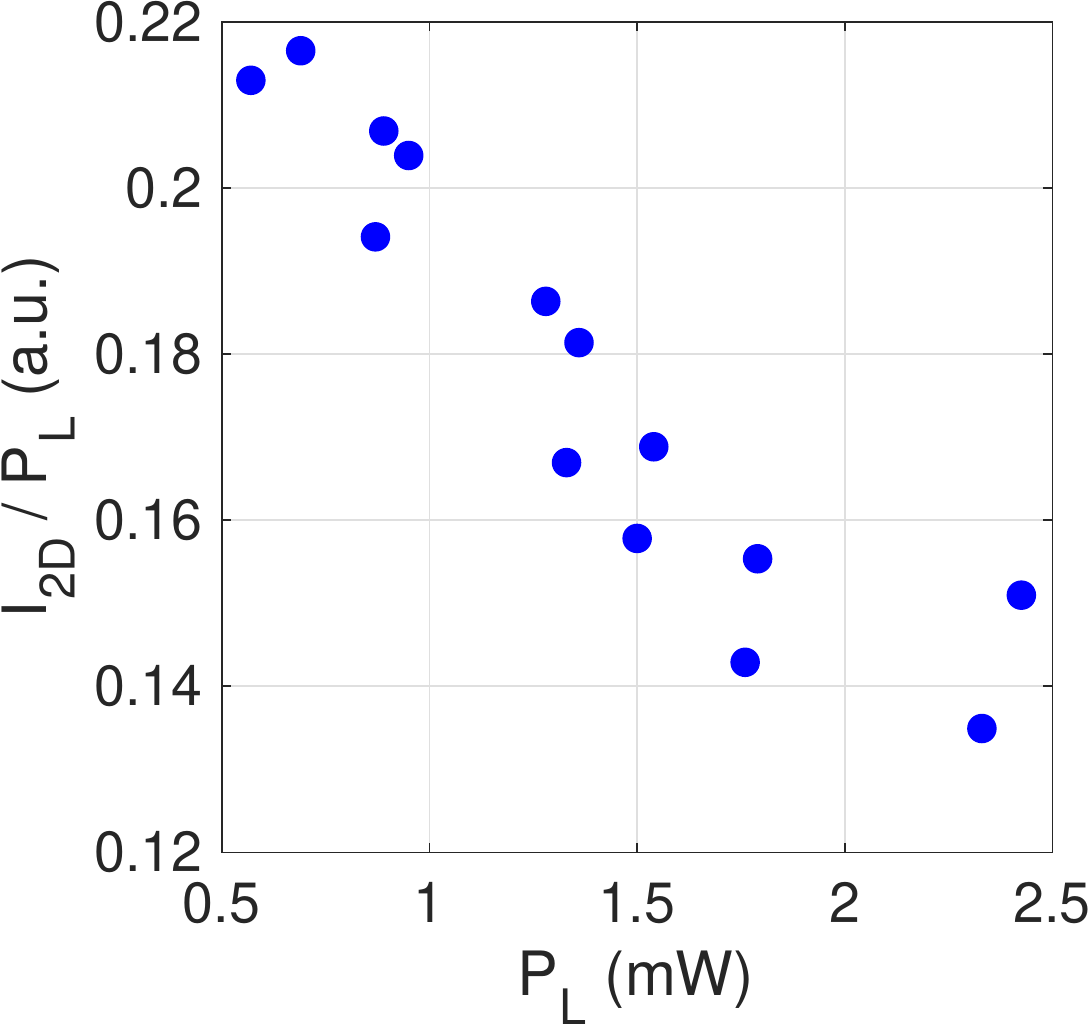}\\
		
		\textbf{c)}&
		\textbf{d)}\\
		
	\end{tabular}
	\caption{Fluence dependence of the G and 2D Raman peaks for a single Layer-CVD graphene. G (a) and 2D (c) peaks at different laser powers ($P_L$). Calculated intensity of the G (b) and 2D (d) peak as a function of laser power. Data are normalized to the incident laser powers.}
	\label{G_2D_review}
\end{figure}

\subsection*{Bandwidth of the G-mode feature}
The Raman lineshape of the G mode represents a convenient way for probing the electronic properties of Gr. Under pulsed excitation the measured G-mode linewidth is also affected by the spectral width of our nearly Fourier transform-limited probe pulses, which is typically $\sim 17$cm$^{-1}$ as stated in the manuscript. 
Specifically, an increase of the electronic temperature $T_e$ is reflected by a broadening of the G band\cite{Ferrante2018} while, on the contrary, the doping induces a spectral narrowing\cite{Froehlicher2015}.
Interestingly, the experimentally detected depletion of the 2D area reported in Fig. 1d) may be in principle rationalized as originating from a charge transfer process, due to an increasing of doping level\cite{BaskoA2D} on the picosecond time scale, with a corresponding G mode narrowing.
In striking contrast, the measured G band spectral line-width, reported in Fig. \ref{figS_fwhmG} for both Gr alone and TMD-Gr vdW heterostructure, show an increase of the bandwidth, which hence has to be ascribed to a pump induced heating.
Moreover, by evaluating the difference between the spectral G bandwidth measured in bare Gr and in coupled WS$_2$-Gr (Fig. \ref{figS_fwhmG}b), an induced extra Raman broadening is observed in the TMD sample. This evidence further rules out the charge transfer interpretation. 

\begin{figure}[h!]
	\centering
	\includegraphics[width=14cm]{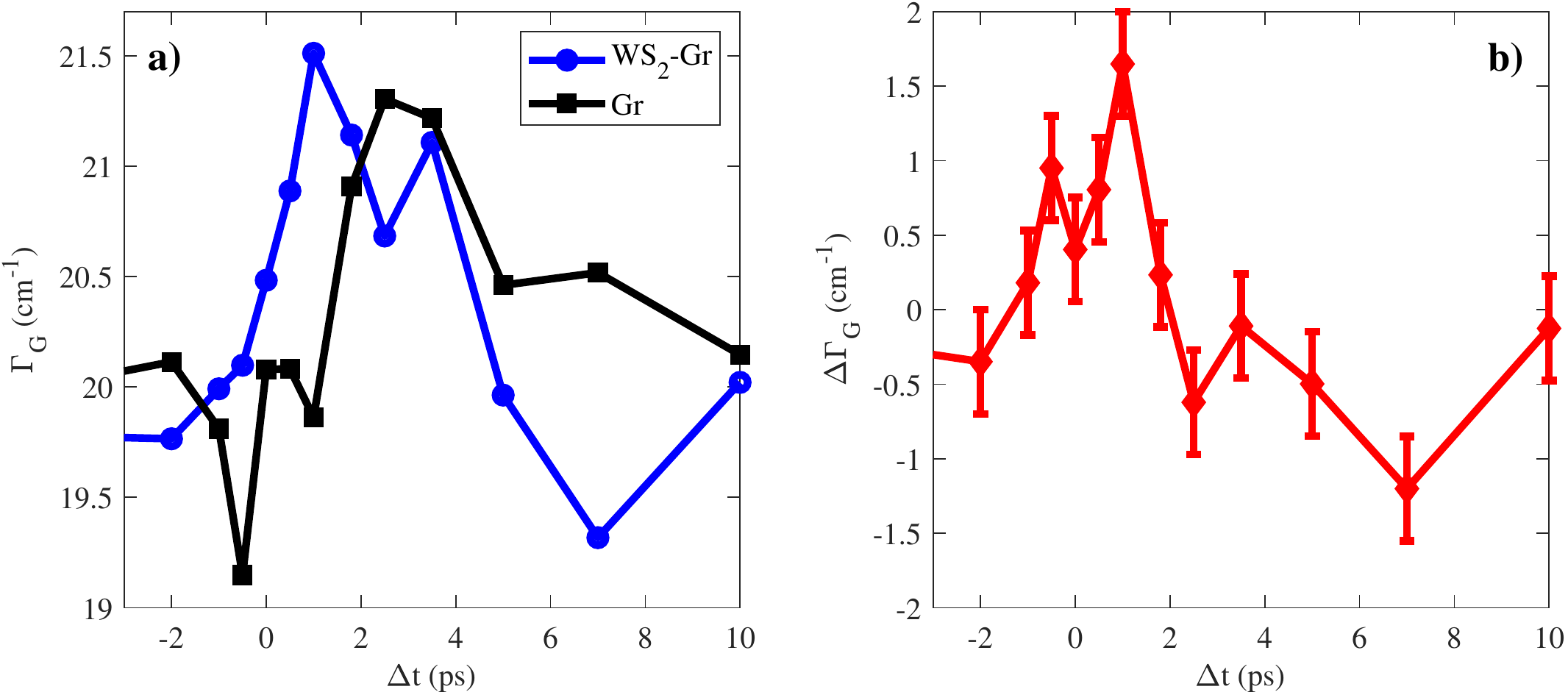}
	\caption{FWHM of G mode of Gr. a) The temporal dependence of the $\Gamma_{\rm G}$ is reported for Gr (black line) and  TMD-Gr vdW heterostructure (blue line). Both the experimental data show an increase of the spectral bandwidth during the overlap between the optical pulses, originated from a  pump pulse induced heating.
		Importantly, as shown by the difference between the spectral G bandwidths reported in panel (b), the broadening in the vdW heterostructure is larger than the Gr alone, testifying a larger heating.}
	\label{figS_fwhmG}
\end{figure}

\section{Frequency of the 2D mode}
In fig. \ref{pos_2d} we show the frequency position of the 2D mode vs the delay time, which is compatible with an effect of the electronic temperature increase, in line with the results of Ref. \cite{Ferrante2018}. This measurement is however at the sensitivity limit of our technique, and further investigations might be needed, to connect the transient evolution of the 2D-mode frequency to the electronic temperature.

\begin{figure}
	\centering
	
	\includegraphics[width=10cm]{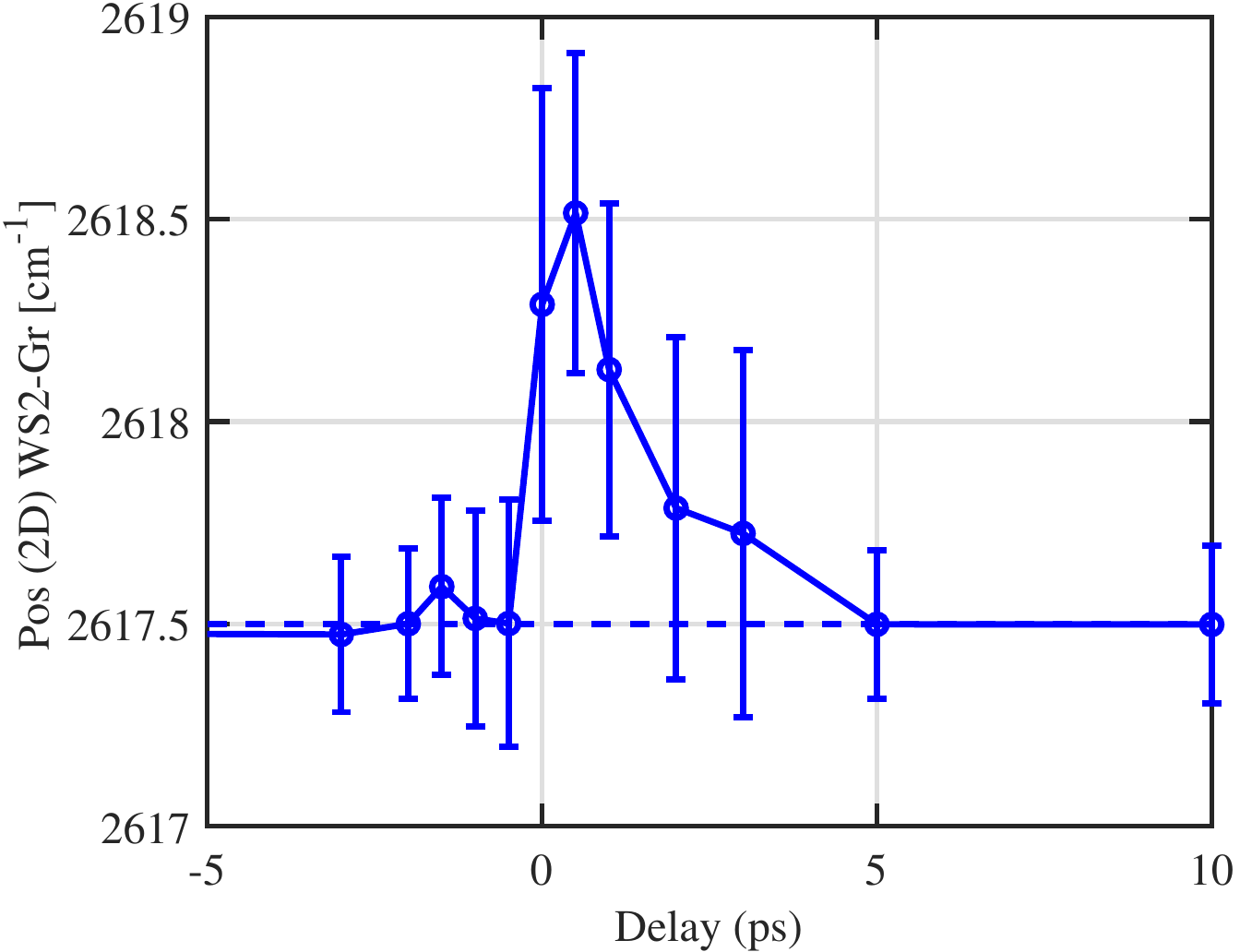}
	
	\caption{The temporal dependence of the position of 2D mode of Gr with the associated 95\% confidence interval.}
	\label{pos_2d}
\end{figure}

\begin{figure}
	\centering
	\includegraphics[width=14cm]{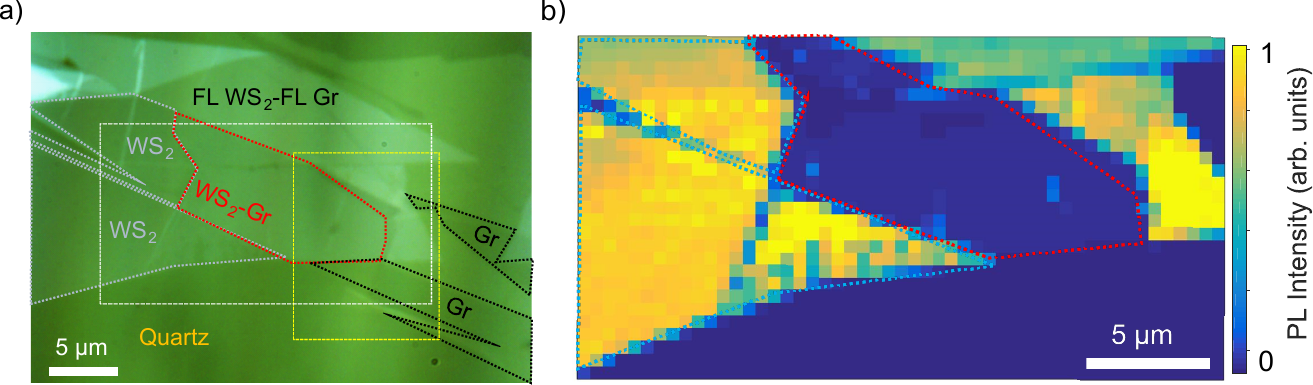} 
	\caption{Photoluminescence characterization in the cw regime.		a) Optical Image of a single layer WS$_2$-single layer graphene van der Waals heterostructure deposited onto a quartz substrate. The heterostructure and nearby graphene are represented with the dashed red and black contours respectively.  FL WS$_2$-FL Gr indicates the region of multilayer 2D materials. The white and yellow rectangles represent the mapped areas for photoluminescence (shown in (b)) and Raman spectroscopy (shown in Fig.~\ref{figS_cwRaman}), respectively. b) PL intensity map of the heterostructure measured in the area limited by the white rectangle in panel (a).
		The spectra were taken in ambient air with a continuous wave 532 nm laser focused onto a $\lesssim1~\mu\mathrm{m}$-diameter spot. The laser power on the sample was 20$\mu$W. }
	
	\label{figS_cwPL}
\end{figure}

\begin{figure}
	\centering
	\includegraphics[width=16cm]{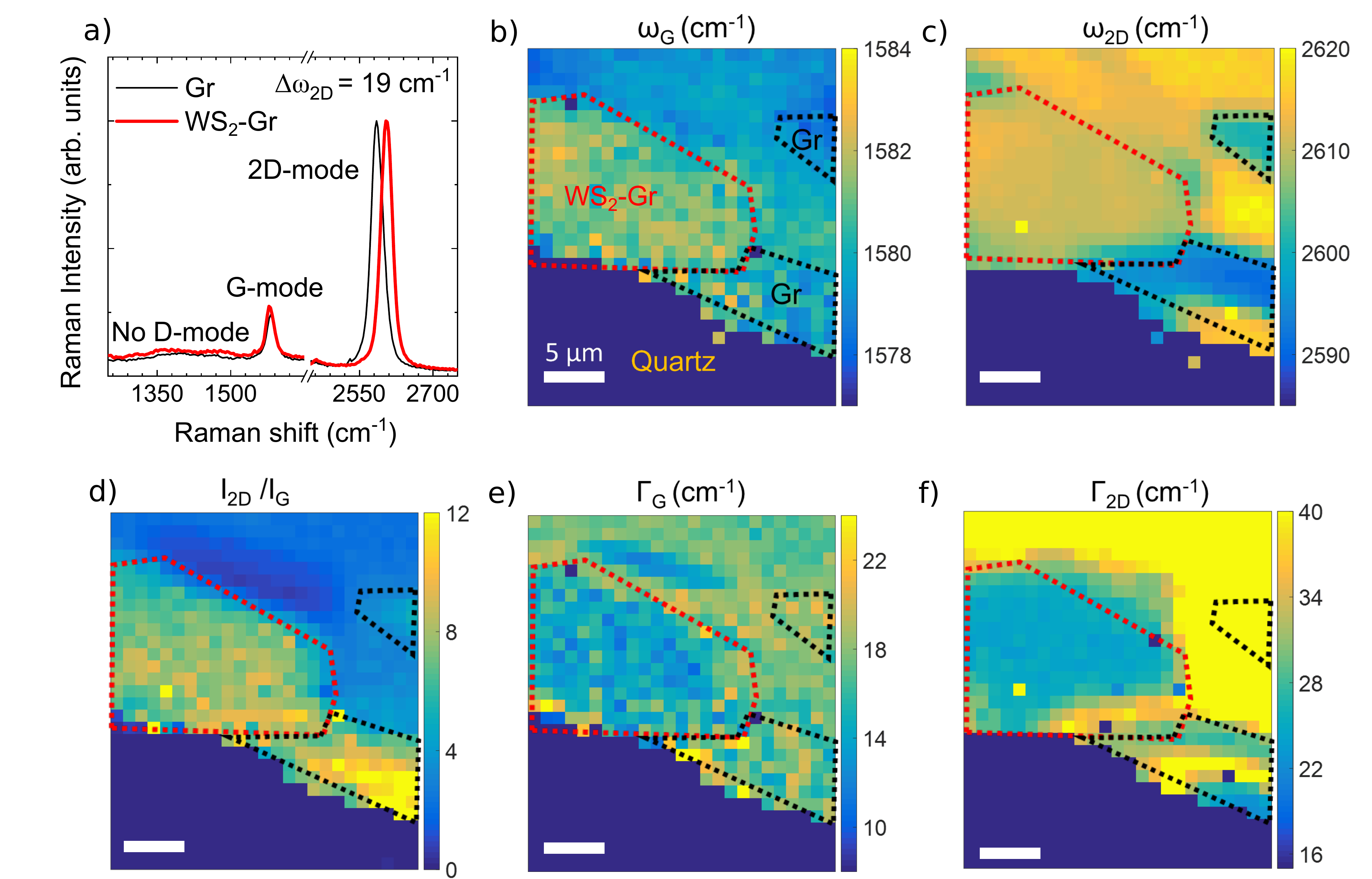} 
	\caption{Raman characterization in the cw regime. a) Raman spectra recorded in the WS$_2$-Gr heterostructure (red contour) and in a nearby graphene monolayer region (black countour). The upper region of the map corresponds to few-layer graphene. b)-f) Hyperspectral Raman maps of the (b) G-mode frequency $\omega_{\rm G}$, (c) 2D-mode frequency $\omega_{\rm 2D}$, (d) ratio between the intensities of the 2D- and G-mode features ($I_{\rm 2D}/I_{\rm G}$), e) G-mode FWHM $\Gamma_{\rm G}$, and (f) 2D-mode FWHM $\Gamma_{\rm 2D}$. All maps have the same scale that for panel (b) and were recorded in ambient air with a continuous wave laser at the same wavelength as for the time-resolved Raman measurements (785~nm). The laser beam is focused onto a $\sim1~\mu\mathrm{m}$-diameter spot. The laser power on the sample was 1.0~mW. As shown in (a) and (c), The 2D-mode feature is upshifted by 19 cm$^{-1}$  on the heterostructure due to dielectric screening as discussed in Refs.~$^{14,72}$. The slight upshift and narrowing of the G-mode feature in the WS$_2$-Gr region as compared to the bare Gr reference is attributed to a larger doping level due to the neutralization of the TMD~$^{6}$ as well  photoinduced charge transfer~$^{14}$. Based on previous studies, we estimate that the doping level in graphene is near $10^{12}~\rm cm^{-2}$, corresponding to a $E_{\rm F}$ of $\le 100~\rm meV$ from the Dirac point in Gr.  }
	
	\label{figS_cwRaman}
\end{figure}

\begin{figure}
	\centering
	\includegraphics[width=10cm]{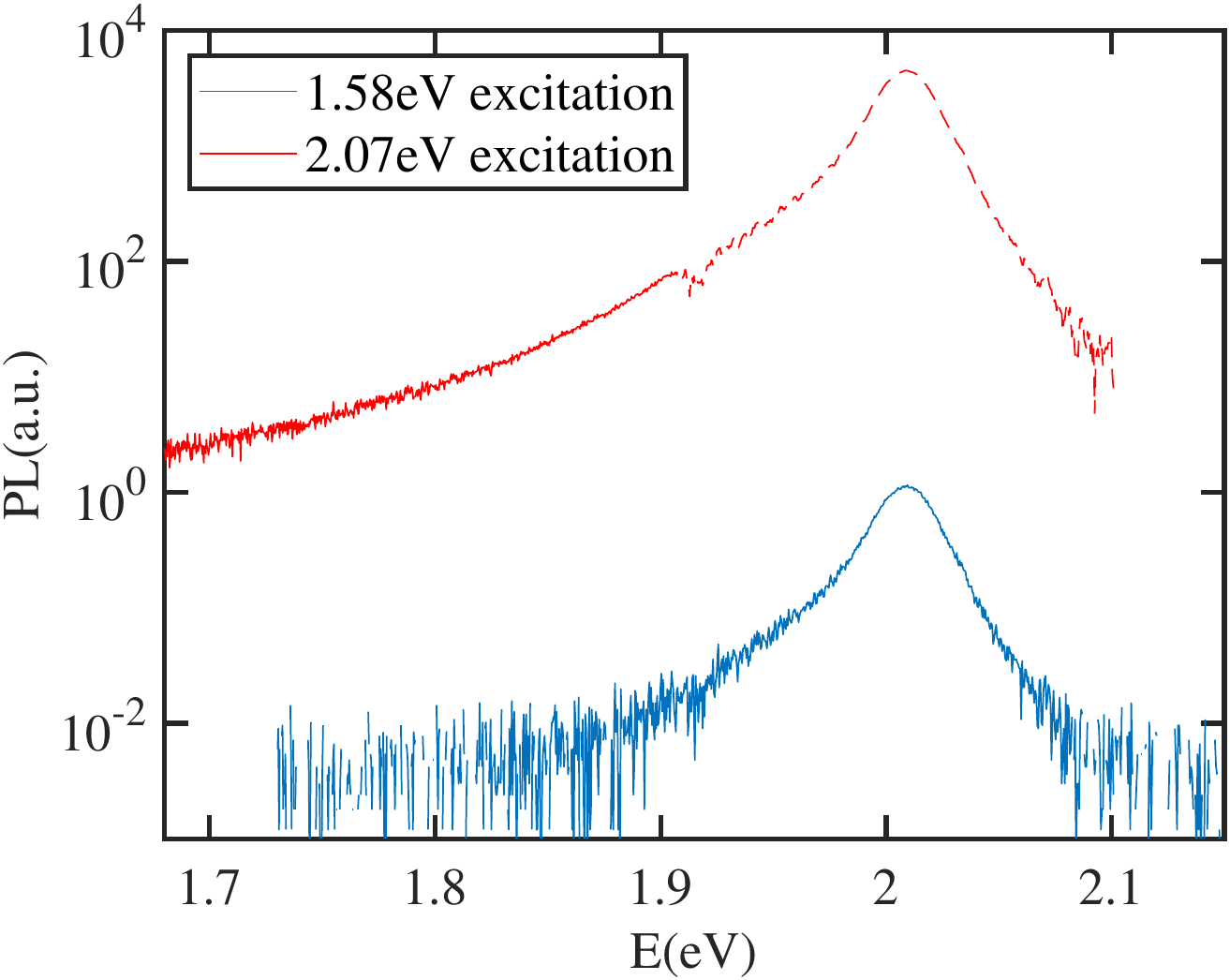} 
	\caption{Photoluminescence at one and two photons. Photoluminescence (PL) obtained exciting WS$_2$ layer at energy above the bandgap  with the pump beam (2.07 eV at 70 $\mu$W) and at energy below the bandgap with the probe (1.58~eV 210~$\mu$W) are reported with red and blue lines, respectively. In the first case a strong one-photon PL is observed. 
		A longpass filter at 650 nm, inserted to suppress the transmission of pump beam, prevent the PL observation at higher photon energy. 
		In the latter case the spectrum of two-photon PL shows a lower efficiency (of a factor $\sim4000$).
		The two photon PL multiplied by a factor 4000 is shown with red dashed line.
	}
	
	\label{figS_PL_twoone}
\end{figure}

\begin{figure}	
	\centering
	\includegraphics[width=14cm]{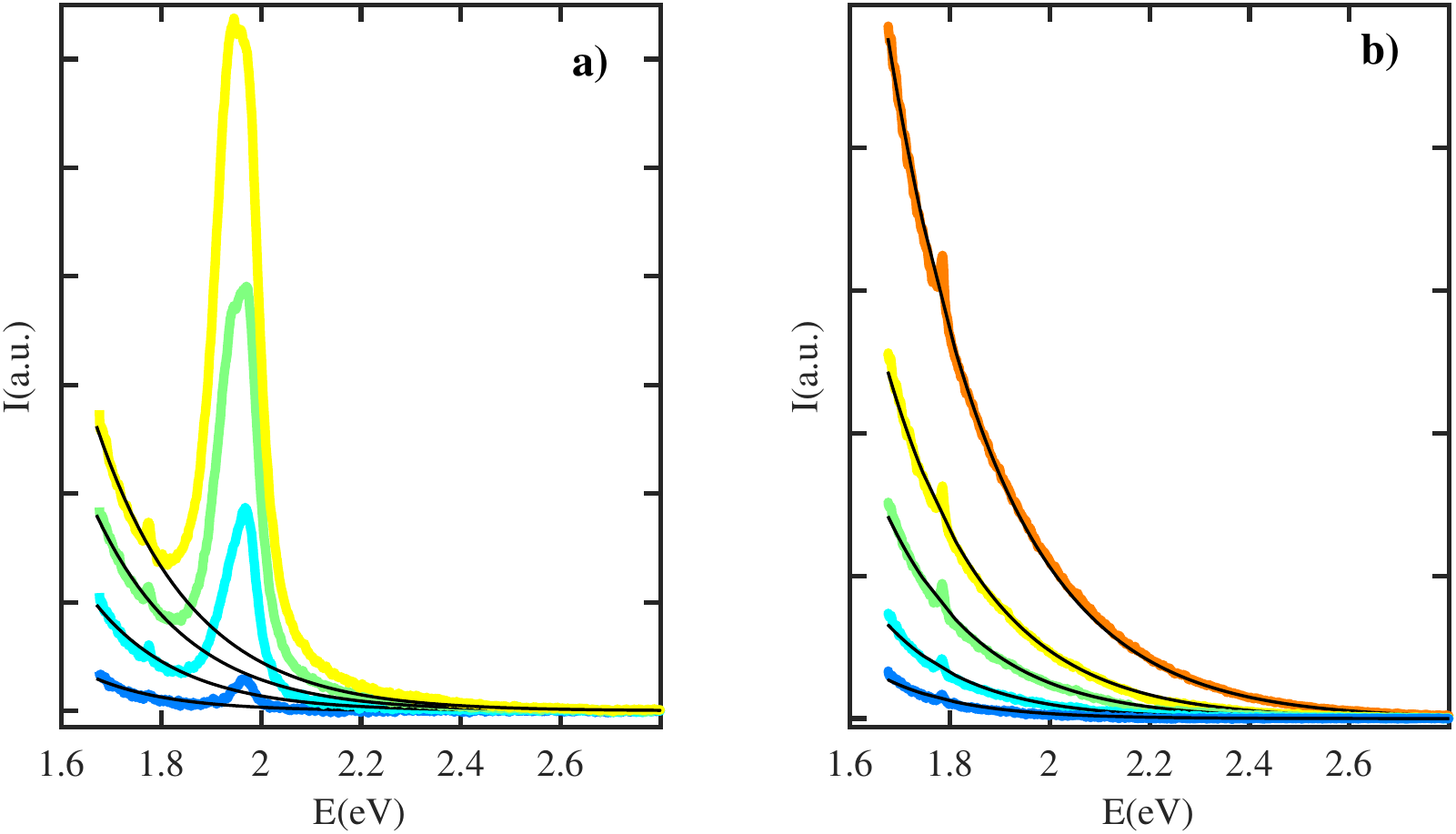} 
	\caption{Hot Luminescence spectra. PRP only experiment at different PRP powers on  WS\textsubscript{2}-Gr a) and  Gr b). Black lines are fits based on Planck's law. The power corresponding to each colour is reported in Fig. \ref{figS_fitHOTPL}. The peak around 1.8eV observed in both figure is the G mode of graphene. The broad peak around 2 eV in a) is the two photon fluorescence of  WS\textsubscript{2}.}
	\label{figS_HOTPL}
	
\end{figure}

\begin{figure}
	\centering
	\includegraphics[width=7cm]{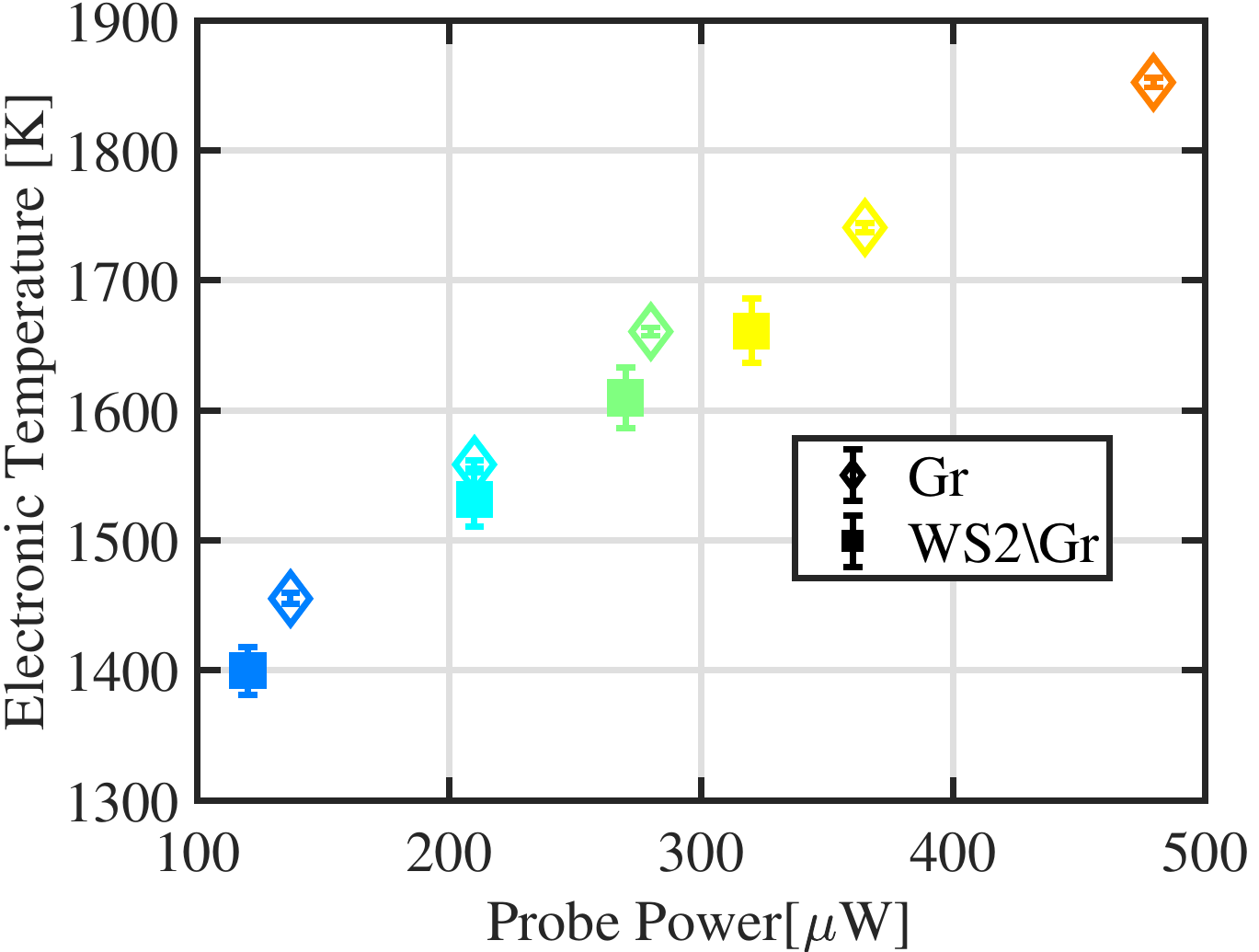} \caption{Light induced $T_e$ in Gr. $T_e$ extracted by the fit of experimental data in Fig. \ref{figS_HOTPL} using Eq. (5). The dependence of $T_e$ on Power is the same for Gr alone (open diamonds) and WS\textsubscript{2}-Gr (full squares).} \label{figS_fitHOTPL}
\end{figure}

\begin{figure}
	\centering
	\includegraphics[width=10cm]{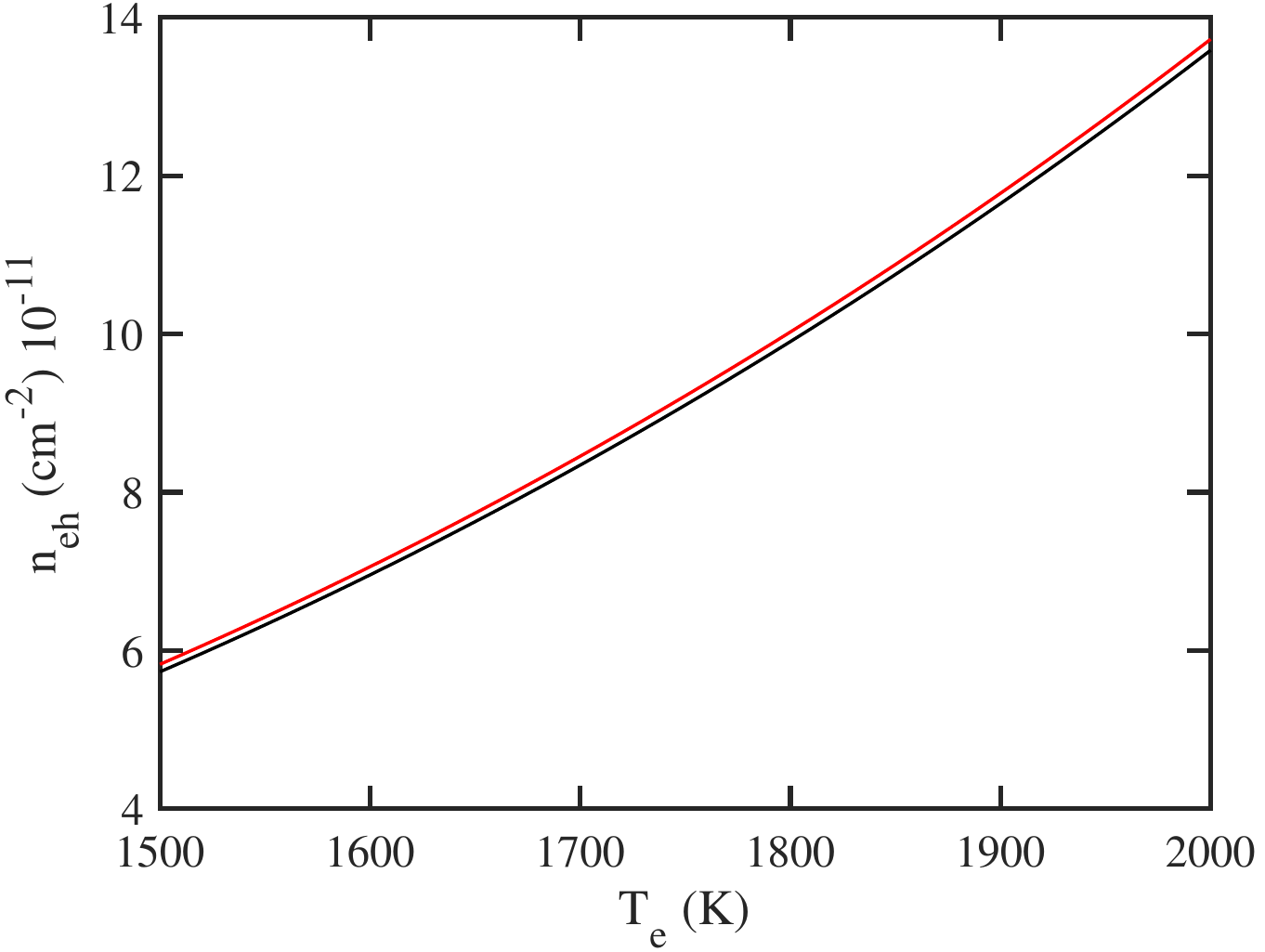} 
	\caption{$n_{\rm eh}$ at different $T_e$ calculated from Eq. (10). Each $n_{\rm eh}$ value represents the number of energy quanta at $E_{\rm PUP}$ necessary for a $T_e$ transition from 0K to the specific temperature reported in the x-axis. The calculation is performed for $E_F=0$ meV and $E_F=30$ meV in black and red lines, respectively.}
	
	\label{figS_Teneh}
\end{figure}

\begin{figure}
	\centering
	
	\includegraphics[width=10cm]{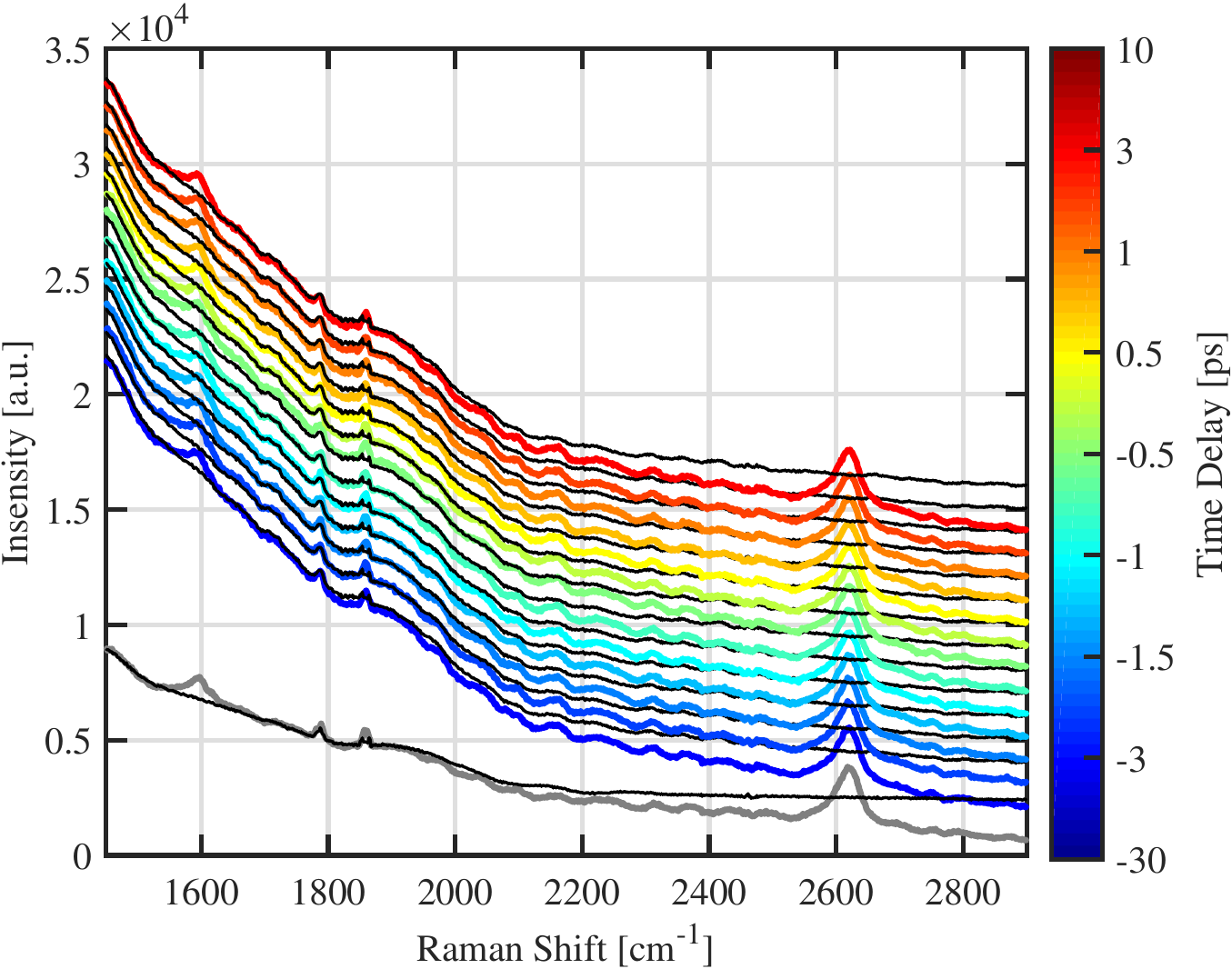}
	
	\caption{Raw Raman spectra. Stacked raw spectra of pump probe experiment on WS\textsubscript{2}-Gr at different time delays (colorbar) and in pump off configuration (gray). The black line below each spectra is the background coming from the substrate, removed in the further steps of the analysis.}
	\label{HeteroSpectrum}
\end{figure} 
\begin{figure}
	\centering
	\includegraphics[width=17cm]{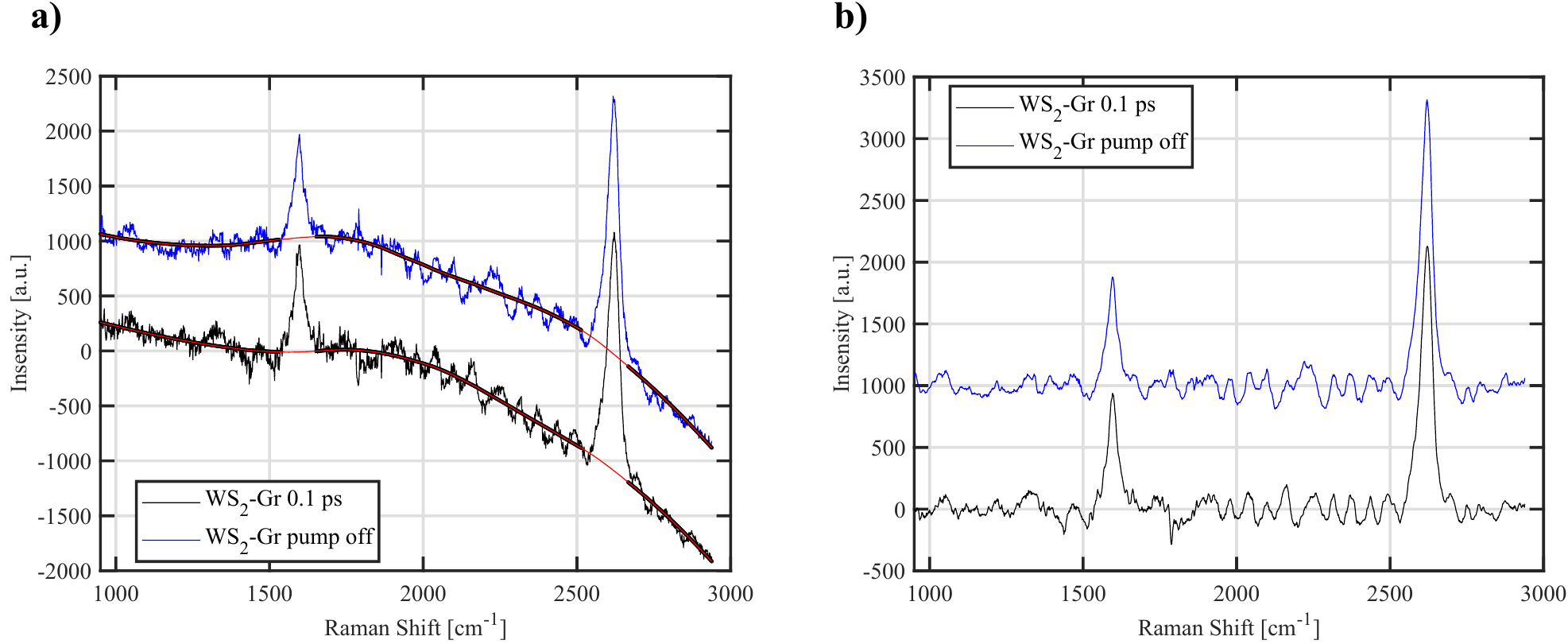}
	
	\caption{Background subtraction. a) Two selected spectra in which the substrate background has been removed. In red the baseline which is subtracted in the next step of the analysis. b) Two selected spectra after the further cleaning procedures.}
	\label{HeteroSpectrum1}
\end{figure}

\begin{figure}
	\centering
	\includegraphics[width=8cm]{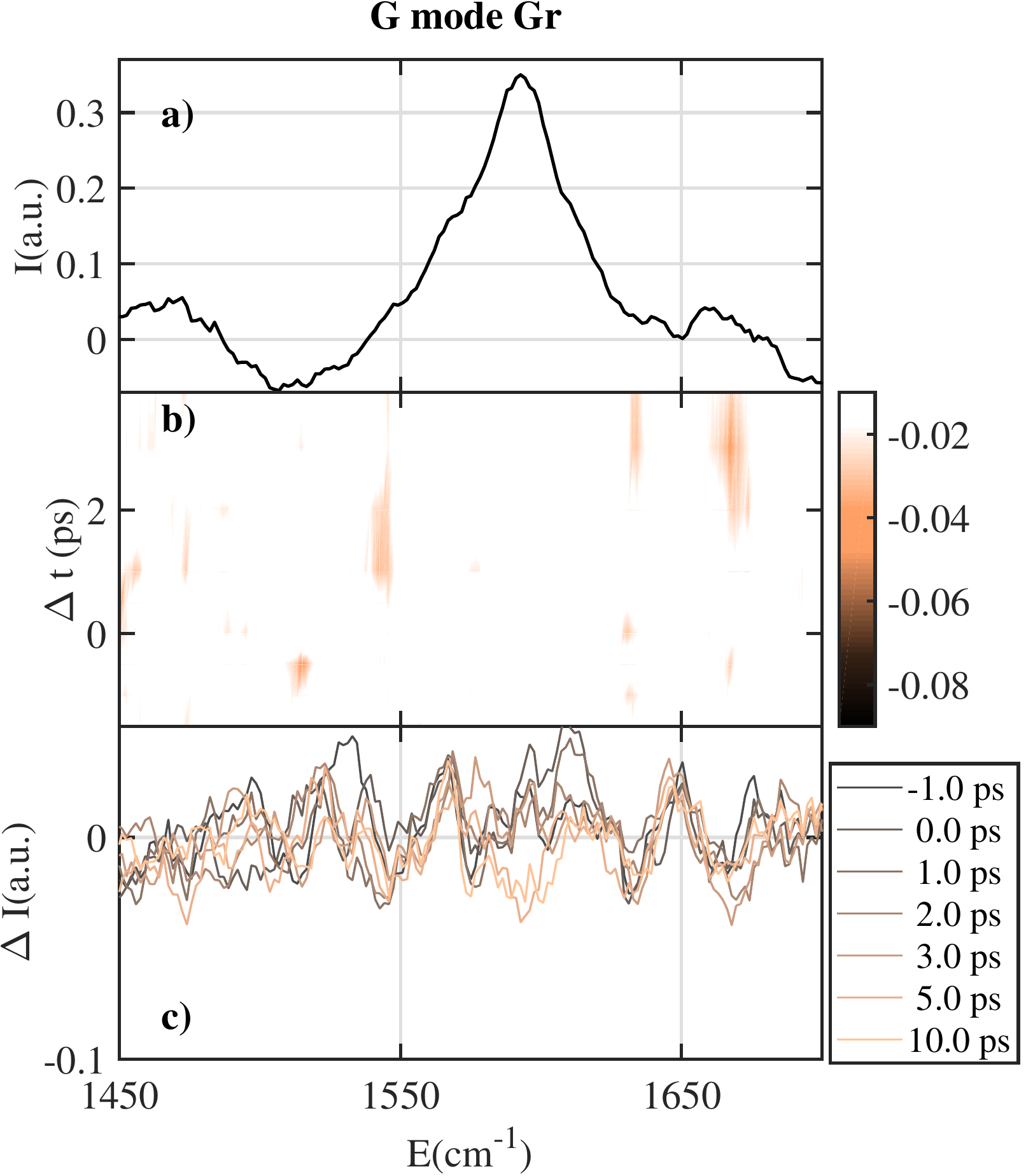} \caption{Temporal profile of G mode in Gr. The Stokes Raman spectra of G (a) mode is reported. The relative differential Raman spectra, calculated as $\Delta I(\Delta t,E)=I(\Delta t,E)-I(-30 ps,E)$ are shown for relative pump-probe time delays in colormaps (b) and spectral plots (c). $I_G$ is  temporally invariant.}
	
	\label{figS_G_gr}
\end{figure}
}

\end{document}